\documentclass[12pt]{article}

\usepackage[english,activeacute]{babel}
\usepackage{natbib}
\usepackage{comment}
\usepackage{float}
\usepackage{hyperref}
\usepackage{mathrsfs}
\usepackage{enumitem}
\usepackage{subfig}
\usepackage[font={small,it}]{caption}
\usepackage{amsmath,amsfonts,amsthm,amssymb}
\usepackage{bm,rotating,multirow,dsfont,graphicx}
\usepackage[usenames, dvipsnames]{color}
\usepackage{url}
\usepackage{multicol}
\usepackage{multirow}
\usepackage[T1]{fontenc}
\usepackage{flafter}
\usepackage{appendix}

\newtheorem{definition}{Definition}

\makeatletter
\def\hlinewd#1{%
\noalign{\ifnum0=`}\fi\hrule \@height #1 %
\futurelet\reserved@a\@xhline}
\makeatother

\def\frac#1#2{{{#1}\over{#2}}}

\def\le{\left}
\def\ri{\right}

\newcommand{\bftab}{\fontseries{b}\selectfont}

\newcommand\simiid{\mathrel{\overset{\makebox[0pt]{\mbox{\normalfont\tiny\sffamily iid}}}{\sim}}}
\newcommand\simind{\mathrel{\overset{\makebox[0pt]{\mbox{\normalfont\tiny\sffamily ind}}}{\sim}}}

\newcommand{\pr}[1]{\textsf{Pr}\left[#1\right]}
\newcommand{\expec}[1]{\textsf{E}\left[#1\right]}

\newcommand{\var}[1]{\textsf{Var}\left[#1\right]}

\newcommand\floor[1]{\lfloor#1\rfloor}

\def\rv{\boldsymbol{r}}

\def\X{\mathbf{X}}\def\Xv{\boldsymbol{X}}\def\xv{\boldsymbol{x}}
\def\Yv{\boldsymbol{Y}}\def\yv{\boldsymbol{y}}

\def\del{\delta}

\def\te{\theta}
\def\vtev{\boldsymbol{\vartheta}}

\def\xiv{\boldsymbol{\xi}}

\def\piv{\boldsymbol{\pi}}

\def\Ga{\Gamma}

\def\Cat{\small{\mathsf{Cat}}}
\def\Dir{\small{\mathsf{Dir}}}

\def\Ber{\small{\mathsf{Ber}}}
\def\Bin{\small{\mathsf{Bin}}}
\def\BetaBin{\small{\mathsf{BetaBin}}}
\def\NegBin{\small{\mathsf{NegBin}}}

\def\Bet{\small{\mathsf{Beta}}}

\def\Geom{\small{\mathsf{Geom}}}

\definecolor{violet}{rgb}{0.56, 0.0, 1.0}
\allowdisplaybreaks

\begin{document}

\def\spacingset#1{\renewcommand{\baselinestretch}{#1}\small\normalsize} \spacingset{1}

\title{A Prior for Record Linkage Based \\
on Allelic Partitions}

\date{}

\author{
 Brenda Betancourt, University of Florida, US\footnote{bbetancourt@ufl.edu} \\ 
 Juan Sosa, Universidad Nacional de Colombia, Colombia\footnote{jcsosam@unal.edu.co}\\
 Abel Rodríguez, University of Washington, US\footnote{abelrod@uw.edu} 
 }

\maketitle

\begin{abstract}
In database management, record linkage aims to identify multiple records that correspond to the same individual. This task can be treated as a clustering problem, in which a latent entity is associated with one or more noisy database records. However, in contrast to traditional clustering applications, a large number of clusters with a few observations per cluster is expected in this context. In this paper, we introduce a new class of prior distributions based on allelic partitions that is specially suited for the small cluster setting of record linkage. Our approach makes it straightforward to introduce prior information about the cluster size distribution at different scales, and naturally enforces sublinear growth of the maximum cluster size -- known as the \emph{microclustering property}. We also introduce a set of novel microclustering conditions in order to impose further constraints on the cluster sizes a priori. We evaluate the performance of our proposed class of priors using simulated data and three official statistics data sets, and show that our models provide competitive results compared to state-of-the-art microclustering models in the record linkage literature. Moreover, we compare the performance of different loss functions for optimal point estimation of the partitions using decision-theoretical based approaches recently proposed in the literature. 
\end{abstract}


\newpage

\spacingset{1.2} 

\section{Introduction}

With the current stream of data, collection and integration of information from multiple sources has become imperative. The process of merging databases and/or removing duplicate records is known as record linkage (RL) \citep{christen-2012}. This is a challenging problem considering that databases often contain corrupted data and lack common unique identifiers across files. Areas of application where RL tasks are prevalent, include public health \citep{gutman-2013, HofRavelliZwinderman17}, human rights \citep{sadinle-2014, sadinle2017, sadinle2018}, official statistics \citep{winkler2014matching, kaplan2018posterior, wortman2019record}, and fraud detection and national security \citep{Vatsalan2017}. 

The seminal work of \citet{fellegi-1969} is the classical reference for a probabilistic approach to identifying links between two files, with a recent extension to three files introduced in \cite{sadinle2013fellegi}. In particular, these approaches rely on record pair similarity weights to determine sets of matches and non-matches. Other work involving the merge of two files includes \citet{belin-1995}, \citet{fienberg-1997}, \citet{larsen-2001}, \citet{tancredi-2011} and \cite{gutman-2013}.  A known caveat of these techniques is that they do not easily generalize to either multiple files or duplicate detection within files. In order to deal with more general scenarios, the RL problem can be viewed as a clustering task in which one or more noisy database records that possibly represent the same latent entity are grouped together. From this point of view, an important feature of RL applications is that, generally, a large number of clusters with a few observations per cluster is expected. From a model-based perspective, popular choices for clustering include finite mixture models and Dirichlet/Pitman-Yor process mixture models (\citealp{muller-2013}, \citealp{casella-2014}, \citealp{miller-2018-mixture}). Although these models have been used in all sorts of applications, including RL \citep{getoor_2006}, they are not well suited for problems with small clusters. Unlike models exhibiting infinitely exchangeable clustering features, models specifically conceived for RL need to generate clusters with a small number of records, even as the size of the data increases \cite{miller-2015}. Within the Bayesian framework, recent advances in latent variable modeling and clustering methods for RL include those of \cite{sadinle-2014}, \cite{steorts-2015-empirical}, \cite{steorts-2015-graphical}. These approaches, however, have the limitation of assuming a uniform prior on the linkage structure which requires strong parameter tuning to achieve sensible RL results. 

In order to formulate more appropriate priors for the small cluster setting of RL, \cite{miller-2015} introduce the concept of \emph{microclustering}, in which the size of the largest cluster of the partition is required to grow sublinearly with the number of records. \cite{betancourt-2016} extended the work of \cite{miller-2015} by introducing a class of Kolchin partition priors (KPPs) for the linkage structure (or cluster assignments) as a way to enforce the microclustering property. However, this formulation is limited by issues of interpretability and identifiability, and also lacks a full characterization of its asymptotic properties. More recently, \cite{betancourt2020random} improved on the weaknesses of the KPP models by proposing a class of prior distributions on random partitions that displays the microclustering property and other desirable characteristics, while preserving computational tractability. 

In this paper, we expand on the existing work of microclustering by proposing a new prior distribution based on allelic partitions. This approach is inspired by the structure of the Ewens’s sampling formula \citep{crane-2016}, which in turn has strong connections with modern Bayesian nonparametric methods. Specifically, allelic partitions are an equivalent representation of partitions which summarizes the number of clusters of each size. In contrast to the previous microclustering approaches, the most appealing feature of this framework for RL applications is being able to handle directly the distribution of the cluster sizes in a natural fashion. Our proposed class of priors is general, however, and can be adapted and used in other microclustering problems \citep{Bloem-Reddy18, klami2016probabilistic}.

The remainder of the paper is organized as follows: Section \ref{sec_model} introduces notation and frames RL as a clustering problem. Section \ref{sec_microclustering} discusses in detail the concept of microclustering, introduces two new microclustering properties that require stronger conditions, and presents a more detailed review of previous work. Section \ref{sec_allelic_partition_prior} discusses our approach based on allelic partitions including inference details. Then, Sections \ref{sec_sims} and \ref{sec_illustrations} explore the performance of our approach compared to the ESC models on five simulated data scenarios and three RL applications, respectively. For the applications, we also explore alternatives for optimal point estimation of the partitions. Finally, we discuss our findings and future work directions in Section \ref{sec_discussion}.

\section{Record Linkage as a clustering task}\label{sec_model}

In this section, we introduce some notation and describe RL from a clustering perspective using a bipartite graph representation of the problem \citep{steorts-2015-graphical}. Consider a collection of $J\geq2$ files. Let $\xv_{i,j}=(x_{i,j,1},\ldots,x_{i,j,L})$ be the attribute data associated with the $i$-th record in file $j$, and let $\X_{j}=[x_{i,j,\ell}]$ be the corresponding $n_j\times L$ array for every $j$. For simplicity, we assume that every record contains $L$ fields in common, field $\ell$ having $D_\ell$ levels. Attribute data of this sort may be considered as either categorical or string-valued but here we focus on a model for categorical data. Let us say, for instance, that data about gender, state of residency, and race regarding $n_j$ individuals in file $j$ are available; in this scenario, $\xv_{i,j}$ is a categorical vector with dimension $L=3$ whose entries have $D_1=2$ (male and female), $D_2=51$ (there are 51 states in the United States including DC), and $D_3=6$ (White, Black or African-American, American Indian or Alaska Native, Asian, Native Hawaiian or Other Pacific Islander, and some other race) levels, respectively. Hence, we can think of records as $L$ dimensional vectors storing attribute information ($L$ fields), while the $j$-th file is composed of $n_j$ records.

Now, let $\yv_k=(y_{k,1},\ldots,y_{k,L})$ be the vector of ``true'' attribute values for the $k$-th latent individual, $k=1,\ldots,K$, where $K$ is the total number of unique individuals in the $J$ files ($K$ could be as small as $1$ if every record in every file refers to the same entity or as large as $n = \sum_j n_j$ if files do not share records at all). Hence, $\Yv=[y_{k,\ell}]$ is an unobserved $K\times L$ attribute matrix whose $k$-th row stores the attribute data associated with the $k$-th latent individual. Next, we define the linkage structure $\xiv=(\xiv_{1},\ldots,\xiv_{J})$, where $\xiv_j=(\xi_{1,j},\ldots,\xi_{n_j,j})$. Here, $\xi_{i,j}$ is an integer from 1 to $K$ indicating which latent individual the $i$-th record in file $j$ refers to, which means that $\xv_{i,j}$ is a possibly-distorted measurement of $\yv_{\xi_{i,j}}$. Such structure unequivocally defines a partition $\mathcal{C}_{\xiv}$ on $\{1,\ldots,n\}$. To see this, notice that by definition, two records $(i,j)$ and $(i^{*},j^{*})$  correspond to the same individual if and only if $\xi_{i,j}=\xi_{i^{*},j^{*}}$. Therefore, $\mathcal{C}_{\xiv}$ is nothing more than a set composed of $K$ disjoint non-empty subsets $\{C_1,\ldots,C_K\}$ such that $\cup_k C_k = \{1,\ldots,n\}$,  where each $C_k$ is defined as the set of all records pointing to latent individual $k$. Hence, the total number of latent individuals $K=K(\xiv)$ is a function of the linkage structure; specifically, $K=\max\{\xi_{i,j}\}$, since without loss of generality we label the cluster assignments with consecutive integers from 1 to $K$. Cluster assignments $\xi_{i,j}$ play a fundamental roll in our approach since they define a linkage structure between files.

\begin{figure}[h!]
	\centering
	\includegraphics[height=0.3\textwidth,height=0.3\textwidth,]{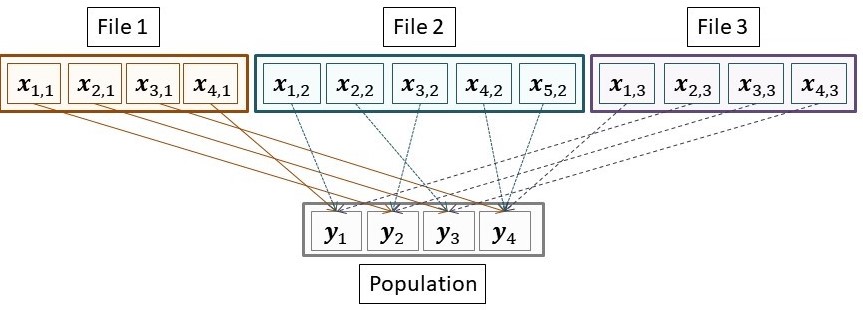}
	\caption{Bipartite graph representation of RL as a clustering task including records $\xv_{i,j}$, latent true attributes $\yv_{k}$, and the linkage structure (edges) $\xiv$.}
	\label{fig_bipartie}
\end{figure}

Figure \ref{fig_bipartie} shows the linkage structure $\xiv$ as a bipartite graph in which each edge links a record to a latent individual. For instance, this figure shows that the sets of records $\xv_{3,1}$, $\xv_{4,2}$, $\xv_{5,2}$ and $\xv_{1,3}$ correspond to the same individual ($\yv_4$). This toy example makes clear that linking records to a hypothesized latent entity is at its core a clustering problem where the main goal is to make inferences about the cluster assignments $\xiv$. In contrast to other clustering tasks, however, we aim to develop an approach that lets the number of records in each cluster be small even for large data sets --  known as \emph{microclusters}, which is characteristic of RL applications \citep{miller-2015, betancourt-2016}. Note that the bipartite graph representation allows for duplicates across and within databases. In practical terms this implies that multiple files can be combined into a single file of size $n = \sum_j n_j$, and we can treat the problem as one of deduplication. Hence, for the remainder of the paper, we drop the file subindex in the notation and simply refer to the attribute data associated with record $i$ as $\xv_{i}$, and the linkage structure as $\xiv=(\xi_1,\ldots,\xi_n)$. 

As far as the linkage structure $\xiv$ is concerned, previous approaches have assumed that every element of $\mathcal{C}_{\xiv}$ is equally likely a priori (i.e. $p(\xiv)\propto 1$), which means that $\xiv$ is restricted to produce partitions composed of equally likely sets of records (singletons, pairs, triplets, etc.) \citep{steorts-2015-graphical}. The uniform prior on $\xiv$ is convenient because it greatly simplifies computation of the posterior. However, such a prior  is not suited for RL tasks since the number of clusters is expected to grow linearly with $n$. For this reason, we devote Sections \ref{sec_ESC} and \ref{sec_allelic_partition_prior} (the latter introduces our proposal) to characterize prior distributions on $\xiv$ that induce the desired behavior for RL tasks.

\section{Microclustering}\label{sec_microclustering}

Finite mixture models and Dirichlet/Pitman-Yor process mixture models are widely used in many clustering applications \citep{miller-2018-mixture}. These models, however, display a sublinear growth of the number of clusters with respect to the number of records. Such a property is unappealing in the context of RL problems because we need to generate a large number of clusters, each with a negligible number of records.  In order to formulate more realistic models for de-duplication, \cite{miller-2015} introduce the \emph{microclustering property}. Formally, the definition states the following:

\begin{definition} \label{def_microclustering}
A random partition $\mathcal{C}_{\xiv}$ of $n$ elements is said to satisfy the microclustering property if $\tfrac{M_n}{n}\xrightarrow{\text{p}}0$ as $n\rightarrow\infty$, where $M_n=\max\le\{|C|:C\in \mathcal{C}_{\xiv}\ri\}$ represents the size of the largest element in $\mathcal{C}_{\xiv}$. 
\end{definition}
That is, the size of the largest cluster in the partition grows sublinearly with $n$, which in turn implies that the number of clusters grows linearly. \cite{miller-2015} and \cite{betancourt-2016} argue that no mixture model can exhibit the microclustering property, unless its parameters are allowed to vary with $n$. In addition, the authors show that in order to obtain nontrivial models exhibiting the microclustering property, we must sacrifice either finite exchangeability or projectivity. In Section \ref{sec_allelic_partition_prior}, we follow their approach by sacrificing projectivity, which seems less restrictive in the RL context. A model for microclustering that sacrifices exchangeability in the context of data with a temporal component is presented in \cite{DiBenedetto2017}. 

Note, however, that Definition \ref{def_microclustering} does not necessarily imply that the size of the largest cluster is finite. Indeed, if for example $\expec{M_n} \sim \mathcal{O}(\log n)$, a simple application of Markov's inequality shows that 
$$
\lim_{n \to \infty} \pr{ \frac{M_n}{n} > \epsilon } \leq \lim_{n \to \infty }\frac{1}{\epsilon} \frac{\expec{ M_n }}{n} = \frac{1}{\epsilon} \lim_{n \to \infty} \frac{\log n}{n}  = 0,
$$
i.e., the microclustering property as initially defined in \cite{miller-2015} is satisfied even though the size of the clusters is allowed to grow unboundedly (both a priori and a posteriori).  Hence, in the sequel we refer to this as the \textit{weak microclustering property}.

In order to impose further constraints on the cluster sizes a priori, we define the \textit{strong microclustering property} as follows:
\begin{definition}\label{def_strong}
A random partition $\mathcal{C}_{\xiv}$ is said to satisfy the strong microclustering property if for any $\epsilon > 0$, there exists finite $M, N>0$ such that $\pr{M_n > M } < \epsilon$ for all $n>N$, where $M_n$ represents the size of the largest element in $\mathcal{C}_{\xiv}$.
\end{definition}

Evidently, the strong microclustering property implies the weak microclustering property (again, by a simple aplication of Markov's inequality), but not viceversa.  However, one shortcoming of this definition is that controlling the size of the largest cluster a priori does not necessarily imply that we have controlled its size a posteriori. In RL applications, where we may have prior information about the size of the clusters, we might want to employ priors that impose stronger constraints. Therefore, we introduce the \textit{bounded microclustering property}: 
\begin{definition}\label{def_strongest}
A random partition $\mathcal{C}_{\xiv}$ of $n$ elements is said to satisfy the bounded microclustering property if, for some constant $M^{*}$, $\pr{M_n > M^*}  = 0$, for all $n$, where $M_n$ represents the size of the largest element in $\mathcal{C}_{\xiv}$.
\end{definition}

By definition, the bounded microclustering property implies both the strong and weak microclustering properties, and ensures that $\pr{M_n > M^{*} \mid \Xv} = 0$. This definition is related to the notion of size-constrained microclustering for finite mixtures discussed in \cite{klami2016probabilistic}, which also assumes that the clusters sizes are bounded in a deterministic fashion. In the remainder of the paper we focus on defining priors that satisfy the bounded microclustering property.

\subsection{Previous Models for Microclustering}\label{sec_ESC}

The work of \cite{betancourt-2016} introduced the idea of Kolchin partition priors (KPPs) as a way to enforce the weak microclustering property  \citep{kolchin1971problem}. This approach consists of placing a prior on the number of clusters, $K \sim \bm{\kappa}$, and then, given $K$, the cluster sizes $S_1,\ldots,S_K$ with $S_k=|C_k|$ are modeled directly as $S_1,\ldots,S_K\mid K \simiid \bm{\mu}$. Here $\bm{\kappa}=(\kappa_s)_{s=1}^\infty$ and $\bm{\mu}=(\mu_s)_{s=1}^\infty$ are probability distributions over  $\mathbb{N}=\{1,2,\ldots\}$. In particular, the authors proposed two models: (a) the NBNB model where both $\bm{\kappa}$ and $\bm{\mu}$ 
belong to the Negative-Binomial family, and a more flexible specification (b) the NBD model where $\bm{\kappa}$ belongs to the Negative-Binomial family and $\bm{\mu}$ is modeled as a random probability vector with a Dirichlet distribution prior. Conditional on $n = \sum_{k=1}^K S_k$, it is straightforward to generate a set of cluster assignments $\xiv=(\xi_1,\ldots,\xi_n)$, which in turn induces a random partition $\mathcal{C}_{\xiv} = \{C_1,\ldots,C_K\}$.

One potential issue with this formulation is that the conditioning on $n$ drastically effects the interpretability of $\bm{\kappa}$ and $\bm{\mu}$, making the elicitation process difficult when information is available a priori. Additional caveats of the KPPs also include a lack of identifiability and of a clear characterization of their asymptotic properties. In order to overcome these limitations, \cite{betancourt2020random} assumes an Exchangeable Sequence of Clusters (ESC) rather than an exchangeable sequence of data points. Under this framework, the prior distribution on a random partition $\mathcal{C}_{\xiv}$ only depends on a distribution over probability distributions $\bm{\mu}=(\mu_s)_{s=1}^\infty$ on the positive integers. In this case, in contrast to the KPPs, $\bm{\mu}$ can actually be interpreted as the distribution of the size of a randomly chosen cluster. It is also important to note that the ESC models satisfy the strong microclustering property when the expectation of $\bm{\mu}$ is finite (i.e.\ $\sum_{s=1}^\infty s\mu_s< \infty$). The authors propose two versions of the ESC model that display a better performance in RL tasks compared to traditional Dirichlet/Pitman-Yor process mixture models: (a) the ESCNB model where $\bm{\mu}=\NegBin(a,q)$; and (b) the ESCD model where $\bm{\mu}=(\mu_s)_{s=1}^\infty$ is modeled as a random distribution with a Dirichlet prior, $\bm{\mu}\sim \Dir(\alpha,\bm{\mu}^{(0)})$, for $\alpha$ is fixed and $\bm{\mu}^{(0)}=\NegBin(a,q)$. In both cases, the parameters $a>0$ and $q\in(0,1)$ are assigned Gamma and Beta priors, respectively. 

In this work, we evaluate the performance of our proposed prior for microclustering,  introduced in Section \ref{sec_allelic_partition_prior}, and compare it to the ESC models using both simulated and real data scenarios (see Sections \ref{sec_sims} and \ref{sec_illustrations}).  In terms of computation, the implementation of the ESC priors is carried out by generating only the first $M^*$ components of $\bm{\mu}$, for a value of  $M^*$ greater than the expected maximum cluster size in the next partition. Hence, from a practical perspective, the ESC priors have a similar flavor to the allelic partition priors that we introduce next.

\section{Allelic partition prior}\label{sec_allelic_partition_prior}

In this section, we introduce a new class of prior distributions on the cluster assignments $\xiv$ based on allelic partitions. Let $\mathcal{C}_{\xiv} = \{C_1,\ldots,C_K\}$ be the partition implicitly represented by $\xiv$ and let $\rv=(r_1,\ldots,r_n)$ be the allelic partition induced by $\mathcal{C}_{\xiv}$, where $r_i$ denotes the number of clusters of size $i$ in $\mathcal{C}_{\xiv}$. For example, the set $\{1, 2, 3\}$ yields five possible partitions: $\{\{1,2,3\}\}$, $\{\{1\},\{2,3\}\}$, $\{\{1,2\},\{3\}\}$, $\{\{1,3\},\{2\}\}$, $\{\{1\},\{2\},\{3\}\}$; which correspond to three possible allelic partitions: $(0,0,1)$, $(1,1,0)$,  $(3,0,0)$. This example makes evident that, in general, each partition $\mathcal{C}_{\xiv}$ corresponds uniquely to an allelic partition $\rv$, but the conversely is not true. Therefore, allelic partitions define equivalence classes on the space of partitions. The notion of allelic partitions will allow us to construct a flexible model for microclustering by assigning appropriate prior distributions on $r_i$. The most appealing feature of this framework for RL applications is being able to explicitly calibrate the maximum cluster size and control the distribution of the cluster sizes. 

Note that, from the definition of allelic partition, it follows directly that $\sum_{i=1}^n i\, r_i = n$ and $\sum_{i=1}^n r_i = K$. Similarly to the KPP models \citep{betancourt-2016}, the construction of the model based on allelic partitions entails conditioning of $n$. However, the limitations that arose in that case from this conditioning are overcome in this context by allowing the parameters of the prior distribution on $r_i$ to vary with $n$ in a natural fashion (see Section \ref{sec_bbap}). To further illustrate the concept of allelic partition, consider the Ewens-Pitman Prior (EPP, \citealp{mccullagh-2006}),
which is intrinsically related to the Dirichlet process.  The probability mass function for the EPP is given by
\begin{equation}\label{eq_CRP_prior}
p(\xiv\mid\theta) = \frac{\Ga(\te)}{\Ga(n+\te)}\, \theta^K \prod_{k=1}^K\Ga(S_k),
\end{equation}
where $\theta$ is an unknown positive parameter. Note that this prior can be factorized as 
\begin{equation}\label{eq_hierarchial_prior_xi}
p(\xiv \mid \theta) = p(\xiv\mid\rv)\,p(\rv\mid \theta),
\end{equation}
where $p(\xiv\mid\rv)= \frac{1}{n!}\prod_{i=1}^n i!^{r_i}\,r_i!\,$ is the uniform distribution on all partitions that belong to the equivalence class represented by $\rv$, and
$$
p(\rv\mid\theta) =\frac{n!}{\theta(\theta+1)\cdots(\theta+n-1)} \prod_{i=1}^{n}\frac{\theta^{r_i}}{i^{r_i}\,r_{i}!} \,,
$$
has support on all possible allelic partitions of the set $\{1, \ldots, n\}$. This representation of the EPP directly motivates the structure of our allelic priors for microclustering. In particular we preserve the same structure for $p(\xiv\mid\rv)$ (which ensures that the prior is finitely exchangeable for any $n$), and replace $p(\rv)$ with a distribution that places its probability on the kind of allelic partitions that are consistent with microclustering applications.

In particular, in the sequel we focus on the bounded microclustering property.  Let $M^* = \max\le\{i\in[n]:r_t >0\text{, for all $t>i$}\ri\}$, $M^* \ll n$, be the size of the largest cluster in $\mathcal{C}_{\xiv}$, i.e., let $M^*$ represent the maximum number of times any one unique record can be repeated in the data set. Our strategy consists in fixing $M^*$ to a reasonable value, and then, placing a distribution on $\rv$ that reflects our prior believes, such that $\pr{r_t = 0} = 1$ for all $t>M^*$. It should be clear that, by fixing $M^*$, this approach satisfies the bounded microclustering property, and consequently the strong and weak properties as well. This type of hard constraint could be of particular practical use in RL scenarios where, due to the data collection mechanism, it is known a priori that there are no duplicates within databases. In that case, the maximum cluster size is expected to be restricted to the number of databases available for deduplication. In cases where there is no strong prior information about the size of the clusters or one wishes to be less restrictive a priori, the value of $M^{*}$ can be chosen to be relatively large to allow for more flexibility (see section \ref{sec_illustrations} for illustrations). Moreover, the number of singletons and the number of latent individuals are easy to calibrate, which is very appealing for RL settings where prior information is available at such a scale.

\subsection{Beta Binomial Allelic Prior (BBAP)} \label{sec_bbap}

In this section, we describe one possible specification of the distribution of the allelic partition for bounded microclustering. In order to specify $p(\rv)$, we first factorize the joint distribution as
$$
p(\rv) =
p(r_{M^*})\,p(r_{M^*-1}\mid r_{M^*})\,p(r_{M^*-2}\mid r_{M^*-1},r_{M^*})\,\ldots\, p(r_1\mid r_{2},\ldots,r_{M^*}).
$$
Moreover,
we assume conditional Binomial distributions for the cluster sizes,
$$
r_{M^*} \sim \Bin(Q_{M^*}, \theta_{M^*}) \quad \text{and} \quad r_{t}\mid r_{t+1},\ldots,r_{M^*}\sim \Bin(Q_t(r_{t+1}, \ldots, r_{M^*}), \theta_t) ,
$$
where the number of trials follow the  recursive specification
$$Q_{M^*} = \floor{n/M^*}\quad\text{and}\quad Q_t(r_{t+1}, \ldots, r_{M^*}) = \floor{(n-\sum_{i=t+1}^{M^*} i\,r_i)/t}\,,$$ 
for $t=2,\ldots,M^*-1$. Finally, $ r_1= n-\sum_{i=2}^{M^*} i\, r_i$ which means that $r_{1}\mid r_2,\ldots,r_{M^*}\sim \delta_{Q_1}$. It is important to note that this particular specification yields cluster size distributions that are consistent with the conditions $\sum_{i=1}^n i\, r_i = n$ and $\sum_{i=1}^n r_i = K$. For instance, for $M^*=2$, the specification of $Q_{M^*}$ respects the restriction that we can at most observe $\floor{n/2}$ clusters of size two in a data set of size $n$.

In addition, the parameters $\theta_t$ controls the proportion of clusters of size $t$ that we expect to observe in the partition. Because of the parameters $\theta_1, \ldots, \theta_{M^*}$ play such a critical role in the model, we increase the versatility of the prior by letting $\theta_t \sim \Bet(a_t,b_t)$, allowing greater control on both the prior mean and the prior variance of each $r_t$. We refer to this prior formulation as the Beta Binomial Allelic Prior (BBAP).

As an example, consider the case of $M^*=2$. Here, it is straightforward to see that the corresponding allelic partition becomes $\rv=(n - 2r_2,r_2,0, 0, \ldots,0)$, which allow us to formulate a hierarchical prior for $\xiv$ only in terms of the number of clusters of size two ($r_2$). Thus, if $M^{*} = 2$ and we denote $a_2 = a$ and $b_2 = b$, we have that
\begin{multline}
	p_{BBAP}(\xiv\mid a, b) = \frac{(n-2r_2)!\,2^{r_2}\, r_2!}{n!} \, \frac{\Ga(\floor{n/2}+1)}{\Ga(r_2+1)\,\Ga(\floor{n/2}-r_2+1)}\\
	\frac{\Ga(r_2+a)\,\Ga(\floor{n/2}-r_2+b)}{\Ga(\floor{n/2}+a+b)}\,\frac{\Ga(a+b)}{\Ga(a)\,\Ga(b)}\,,
\end{multline}
with the expected number of singletons a priori being
$$
\expec{ r_1 } = n - 2 \left \lfloor  \frac{ n }{ 2 } \right \rfloor \frac{a}{a + b} \approx \frac{b n}{a+ b}\,,
$$
with variance
$$
\var{ r_1 } = 4 \left \lfloor  \frac{ n }{ 2 } \right \rfloor \left( a + b + \left \lfloor  \frac{ n }{ 2 } \right \rfloor \right) \frac{ab}{(a + b)^2(a+b+1)}\,.
$$
As we discussed before, the number of singletons is one of the quantities for which there is often strong prior information in RL problems. Therefore, these expressions are key for prior calibration. In fact, more generally
$$
\expec{r_{M^*}} = \frac{a_{M^*}}{a_{M^*} + b_{M^*}} \left\lfloor \frac{n}{M^*} \right\rfloor = \frac{a_{M^*}}{a_{M^*} + b_{M^*}} Q_{M^*}$$
and
$$
\expec{ r_t } = \frac{a_t}{a_t + b_t} \sum_{s_{t+1}=0}^{Q_{t+1}} \cdots
%
%
\sum_{s_{M^*}=0}^{Q_{M^*}}
Q_t
q\left(s_{t+1}, \ldots,s_{M^*}\right)  ,
$$
where
\begin{multline}
q\left(s_{t+1}, \ldots,s_{M^*}\right) = \BetaBin(s_{M^*}\mid Q_{M^*}, a_{M^*}, b_{M^*})
\prod_{k=t+1}^{{M^*}-1} \BetaBin(s_k \mid Q_k, a_{k}, b_{k})
\end{multline}
for $Q_k \equiv Q_k(s_{k+1}, \ldots, s_{M^*})$ and $t=2,\ldots,M^{*}-1$. These expressions, however, are too convoluted to be of real practical utility. In the following section, we provide some practical guidelines to calibrate the hyperparameters of the model to prior knowledge.

\subsection{BBAP Calibration}  \label{sec_calibrate}

In general, for RL applications where the percentage of duplication is low, we would like $\theta_t$ to decrease fast with $t$ to reflect the fact that we expect most items to be singletons. On the other hand, when attempting to combine $J$ files in which we expect substantial overlap, we would typically pick $M^* \ge J$ and use relatively large values of $\theta_J$. For example, in the case $M^*=2$, given a prior probability of duplication $\pi$ (often less than 0.3 in many deduplication settings) along with a corresponding coefficient of variation $\gamma$ (e.g., $\gamma=0.5$ for vague levels of precision), it is straightforward to see that by letting 
$$
a_2=\frac{1-\pi(1-\gamma^2)}{\gamma^2} \quad \text{and} \quad b_2=a_2\frac{(1-\pi)}{\pi},
$$
we obtain the desired prior calibration. For $M^{*} > 2$, a similar procedure can be implemented using numerical computations that leverage the recursive nature of the prior. More specifically, after providing a vector of prior probabilities for the cluster sizes $\piv=(\pi_2,\ldots,\pi_{M^*})$ based on prior knowledge, the elicitation of the hyperparameters $a_t$ and $b_t$ can be done recursively according to the coefficient of variation chosen by the practitioner.

Considering that many RL applications display a distribution of cluster sizes with a `geometric like' decay (i.e. a large number of singleton clusters is expected), we explore a default calibration of the BBAP that exhibits this behavior. The prior is calibrated assuming values for the prior probabilities of the clusters of each size from a truncated Geometric distribution, $\piv=\Geom{(p)}$. We also consider a truncated Negative Binomial, $\piv=\NegBin{(r,p)}$ with the purpose of assessing the sensitivity of the results to the prior calibration. Furthermore, in cases where the data collection mechanism naturally informs the maximum cluster size, for example merging $J$ databases known to have no duplication within, we can choose $M^{*}=J$ to obtain sensible RL results. When there is no strong prior information about the size of the clusters or one wishes to be less restrictive a priori, the value of $M^{*}$ can be chosen to be relatively large to allow the maximum cluster size to be estimated from the data without risk of truncation a priori. See Sections \ref{sec_sims} and \ref{sec_illustrations} for illustrations of these different calibrations and their effects on posterior inference.  

\subsection{Posterior Inference for BBAP model}\label{sec_computation}

In order to obtain samples from the BBAP model a posteriori, we derive the probability distribution of a record being assigned to an existing or new cluster conditional on the current partition of the data and the prior parameters. This type of assignment rule has been widely used in the context of Dirichlet/Pitman-Yor processes and it is especially useful for computational tractability in sampling of random partitions. For non-projective models like the BBAP model, we refer to these cluster assignment probabilities as \emph{reallocation probabilities}.  Given the conditional EPPF in equation (\ref{eq_hierarchial_prior_xi}) and that
$$
p(\xi_i\mid\xiv_{-i},\rv) = \frac{p(\xiv\mid\rv)}{p(\xiv_{-i}\mid\rv_{-i})}\, \frac{p(\rv)}{p(\rv_{-i})},
$$
the reallocation probabilities for the BBAP model are  given by
\begin{equation}\label{eq:prediction_rule}
p(\xi_i=k \mid \xiv_{-i},\rv_{-i})
\propto
\left\{
\begin{array}{ll}
(|k|+1) \,
\dfrac{r_{-i,|k|+1} + 1}{r_{-i,|k|}} \, \dfrac{p(\rv)}{p(\rv_{-i})} & \hbox{if }k=1,\dots,K_{-i},
\\
(r_{-i,1} + 1) \, \dfrac{p(\rv)}{p(\rv_{-i})} & \hbox{if }k=K_{-i}+1\,,
\end{array}
\right.
\end{equation}
where $|k|=1,\ldots,M^*-1$ is the size of cluster $k$, and $r_{-i,|k|}$ and $K_{-i}$ are the number of clusters of size $|k|$  and the total number of clusters in $\mathcal{C}_{\xiv} \setminus i$, respectively. While the term $p(\rv)/p(\rv_{-i})$ can be readily simplified, its evaluation is straightforward and has a low computational cost. 

Posterior inference using the BBAP is performed by introducing the corresponding likelihood terms of the RL model in the reallocation probabilities. Given that standard Gibbs sampling algorithms are too slow for large data sets with many small clusters, we utilize a modified version of the Chaperones Algorithm initially proposed in \cite{miller-2015} to obtain samples from the full conditional distribution of $\xiv$.  The Chaperones algorithm is similar in spirit to existing split–merge Markov chain sampling algorithms \citep{jain2004split} but exhibits better mixing properties in microclustering settings. The modified version that we implement accelerates the convergence of the algorithm by using a non-uniform proposal to select the `chaperone records' that favors records with common field values while still assigning probabilities greater than zero to all possible record pairs \citet{betancourt2020random}. In the following section, we describe a specific RL model formulation used to illustrate our prior proposal. Note, however, that our allelic partition approach is general and can be used with other RL models or  adapted to other microclustering applications beyond RL. 

\subsubsection{Record Linkage Model} \label{sec_rlmodel}

For the simulations and applications presented in the remainder of the paper, we follow the RL model proposed by \cite{steorts-2015-graphical}. Here, each field is modeled depending on whether it is distorted or not.  If $x_{i,\ell}$ is not distorted,  that particular field is left intact by giving it a point mass distribution at the true value; otherwise, a categorical (multinomial) distribution is placed over all the categories of that particular field. In summary, assuming that the attribute data $x_{i,\ell}$ are conditionally independent given the cluster assignments $\xi_{i}$ and the true population attributes $y_{n,\ell}$, we have that:
\begin{equation}\label{eq_model_RL_part_2}
x_{i,\ell}\mid y_{\xi_{i},\ell},w_{i,\ell},\vtev_{\ell}\simind \left\{
\begin{array}{ll}
\del_{y_{\xi_{i},\ell}},        & \hbox{$w_{i,\ell}=0$;} \\
\Cat\le( \vtev_{\ell} \ri), & \hbox{$w_{i,\ell}=1$,} \\
\end{array}
\right.
\end{equation}
where $\delta_a$ is the distribution of a point mass at $a$, $w_{i,\ell}$ are distortion indicators, and $\vtev_{\ell}$ is a $D_\ell$-dimensional vector of multinomial probabilities. We simply let $w_{i,\ell}\mid\psi_\ell\simind \Ber(\psi_\ell)$ where $\psi_\ell$ represents the distortion probabilities of the fields, and fix $\vtev_{\ell}$ at the empirical distribution of the data.  By integrating $w_{i,\ell}$ out, the likelihood in equation (\ref{eq_model_RL_part_2}) is now:

\begin{equation}\label{eq_model_RL_part_2f}
x_{i,\ell}\mid y_{\xi_{i},\ell},\psi_\ell,\vtev_{\ell} \simind (1-\psi_\ell)\del_{y_{\xi_{i},\ell}} + \psi_\ell \vtev_{\ell}.
\end{equation}
 
In order to complete the model specification, we let $y_{k,\ell}\mid\vtev_{\ell} \simind \Cat(\vtev_{\ell})$ and assign independent priors for the distortion probabilities of the fields, $\psi_\ell  \simind \Bet(c_\ell,d_\ell)$. Finally,  we utilize the ESC and BBAP microclustering priors for the linkage structure $\xiv$.  The distortion parameters $\psi_\ell$ capture the noise of the data  and their values are expected to remain small (usually below 10\%) to obtain sensible RL results.

\section{Simulation Study}\label{sec_sims}

In this section, we explore the behavior of the proposed BBAP models, compared to the ESC models, in five different simulation scenarios. These scenarios are chosen to explore the flexibility of the microclustering priors and their ability to recover the true partition beyond the conventional `geometric like'  behavior assumed in many RL applications. In order to evaluate the sensitivity of the results to prior calibration, we use a default Geometric calibration, $\piv=\Geom{(p)}$ with $p=0.5$ -- denoted as BBAPG, as well as a truncated Negative-Binomial specification, $\piv=\NegBin(r,p)$ with $r=4$ and $p=0.5$ -- denoted as BBAPNB. The parameter values of the Negative-Binomial specification reflect a prior mode for the distribution of cluster sizes between 2 and 3. In both cases, we use a coefficient of variation of $\gamma=0.25$ to reflect relatively vague levels of precision in the calibrations  (recall the discussion in Section \ref{sec_calibrate}). For the ESC models, we set $\alpha = 1$,  $a \sim \text{Gamma}\left(1, 1\right)$, and  $q \sim \text{Beta}\left(2, 2\right)$. These values have been previously suggested as defaults and shown to work well \citep{betancourt2020random}. For computational and comparison purposes, we work with a truncated version of the ESC models in which only the first $M^*$ components of $\bm{\mu}$ are generated. 

\begin{figure}[h!]
\includegraphics[width=\textwidth]{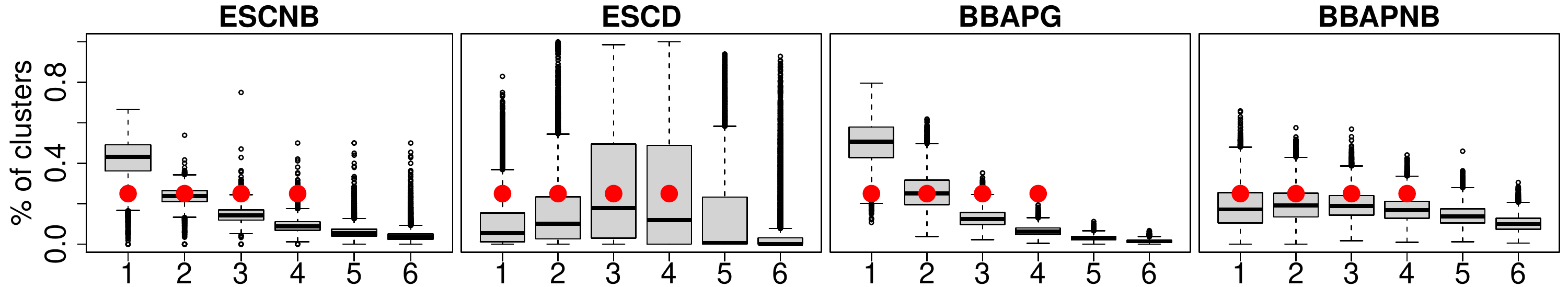}
\includegraphics[width=\textwidth]{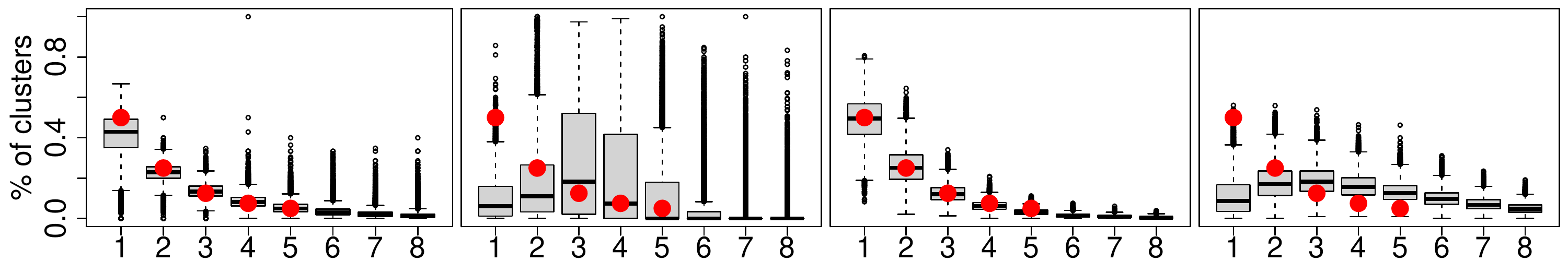}
\includegraphics[width=\textwidth]{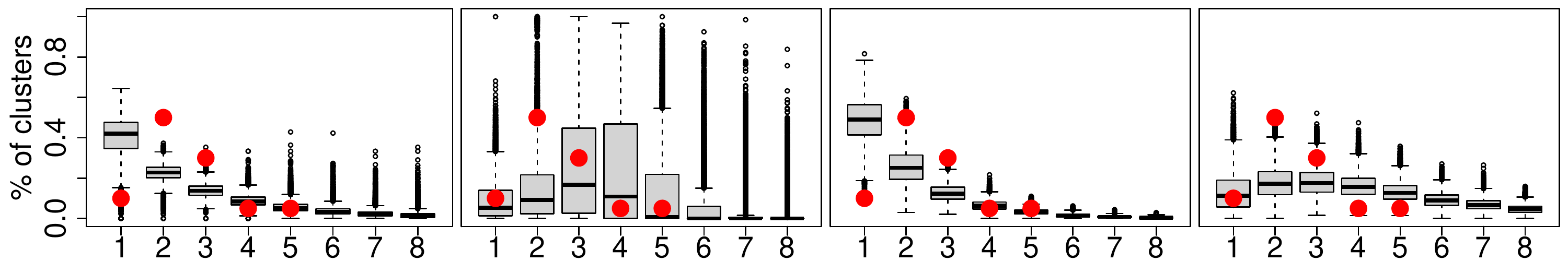}
\includegraphics[width=\textwidth]{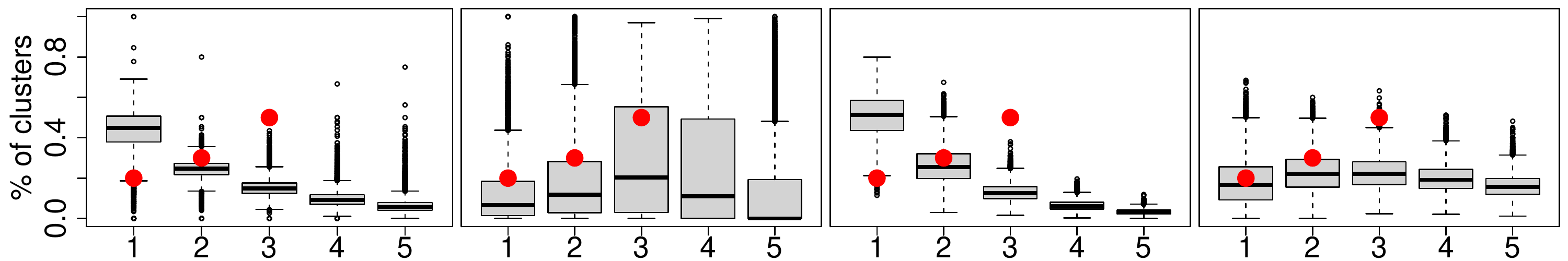}
\includegraphics[width=\textwidth]{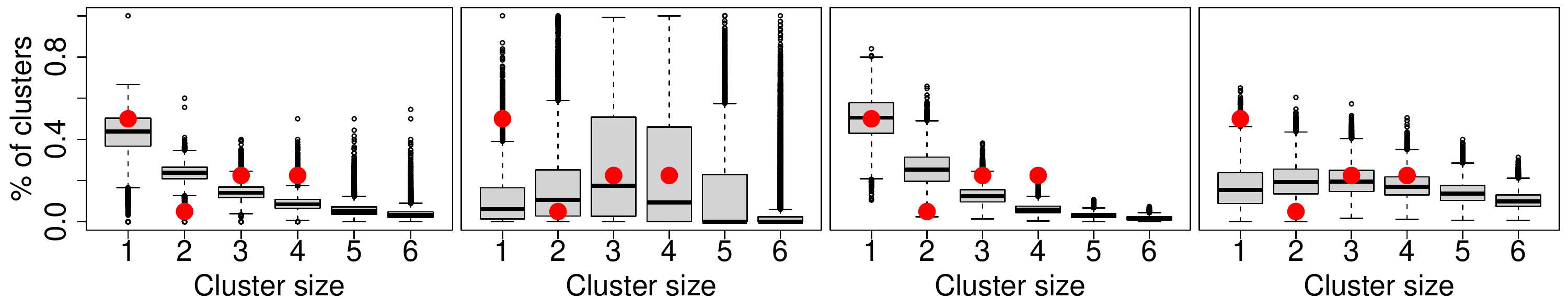}
\caption{Prior distribution of the allelic partition (boxplots) for ESC and BBAP models, and true data partition (red dots) for five simulated scenarios with $K=200$ clusters and distortion probability of the fields $\psi_\ell = 0.05$ (rows).} \label{fig:prior_samples_sims}
\end{figure}

For the simulation, we generate ten data sets obtained from the combination of five different partitions of $K=200$ clusters and two fixed values of the distortion probabilities of the fields -- $\psi_\ell = 0.01$ and $0.05$. The RL task is expected to become more challenging for higher levels of distortion of the fields. In all cases, we set $M^*$ to be one and a half times the true maximum cluster size to generate the prior samples.  Figure \ref{fig:prior_samples_sims} displays the true data partitions and prior samples from the two BBAP and ESC models for the data sets simulated with $\psi_\ell = 0.05$ (similar behavior is observed for $\psi_\ell = 0.01$).  As Figure \ref{fig:prior_samples_sims} shows, scenarios (rows) 1 and 2 display uniform and geometric behavior of the partition, respectively, while scenarios 3 and 4 are more unconventional in that the proportion of singleton clusters is low compared to other cluster sizes. Finally, the  cluster size distribution of the last scenario can be thought of as a mixture of the previous ones. The behavior of ESCNB and BBAPG in terms of the number of clusters of each size is quite similar, although the rate of decay for BBAPG seems to be faster. Furthermore, the behavior of the ESCD prior is quite different from that of the alternatives. In particular, ESCD induces very skewed marginal priors for the proportion of clusters of any given size, and favors configurations in which the most frequent cluster size is between 3 and 4. Finally, the BBAPNB prior distributes the proportion of clusters of any given size more evenly but favors cluster sizes between 2 and 3 as expected from the parameter values used for the calibration ($r=4$ and $p=0.5$).

\begin{figure}[h!]
\includegraphics[width=\textwidth]{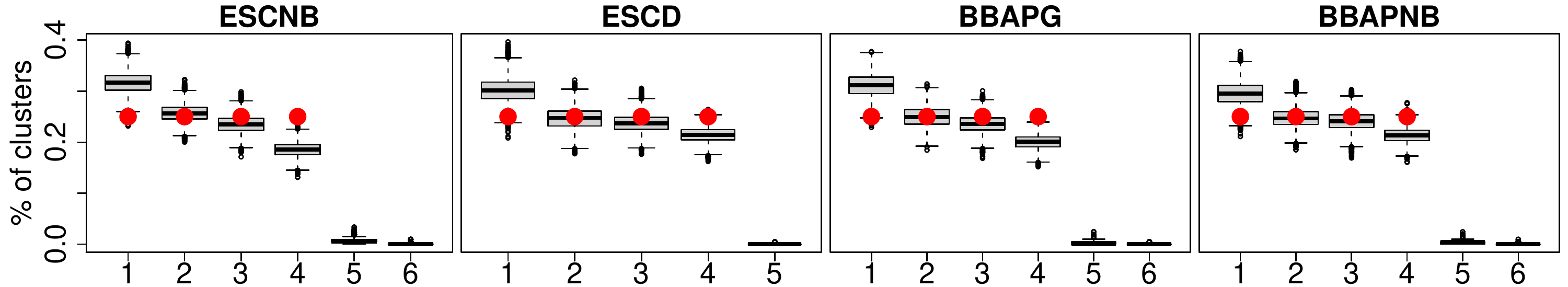}
\includegraphics[width=\textwidth]{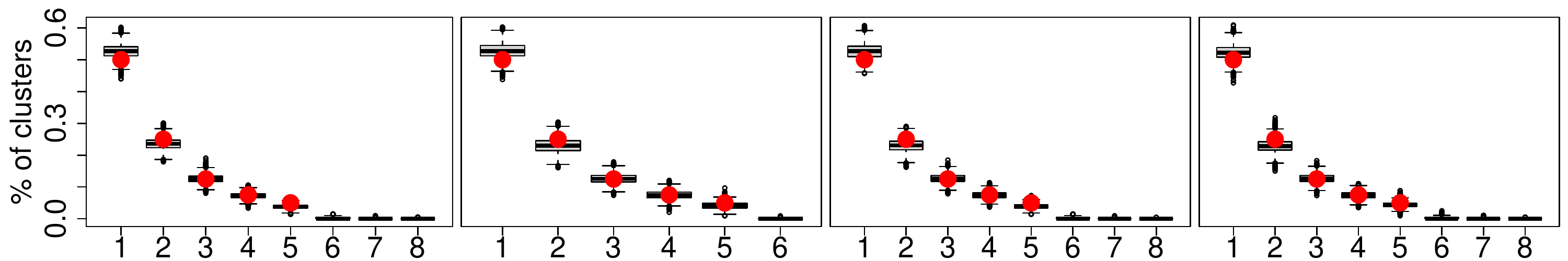}
\includegraphics[width=\textwidth]{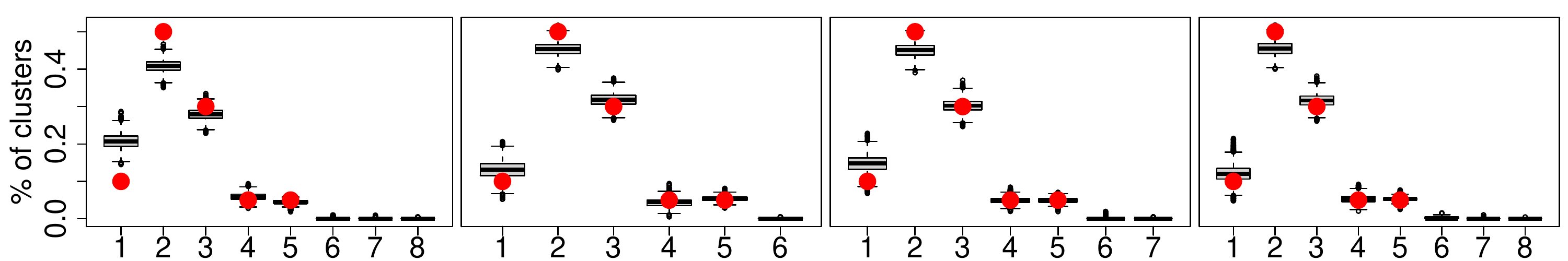}
\includegraphics[width=\textwidth]{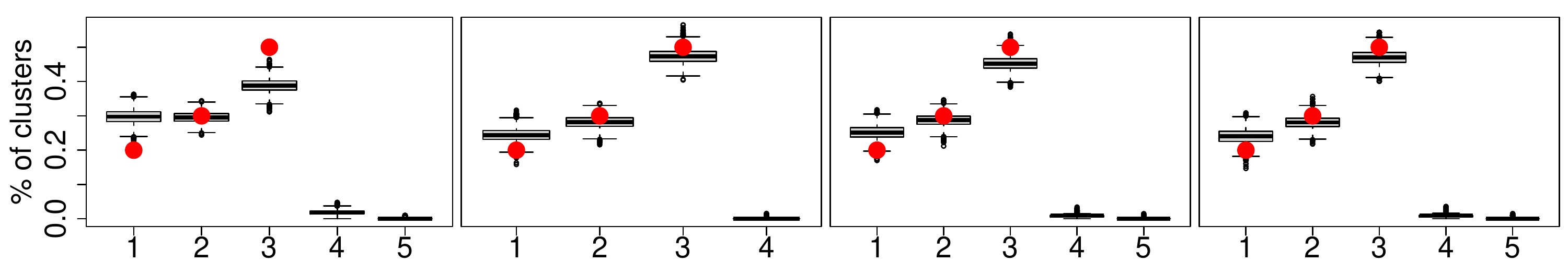}
\includegraphics[width=\textwidth]{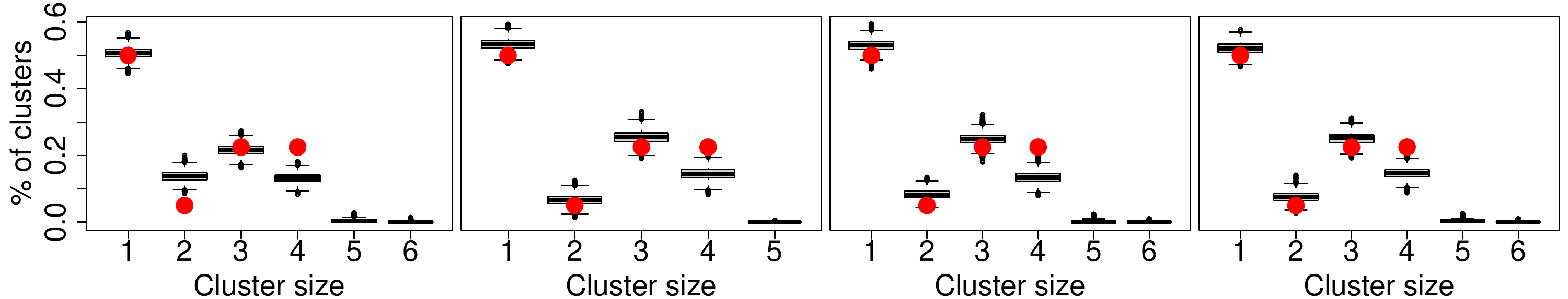}
\caption{Posterior distribution of the allelic partition (boxplots) for ESC and BBAP models, and true data partition (red dots) for five simulated scenarios (rows) with $K=200$ clusters and distortion probability of the fields $\psi_\ell = 0.05$.} \label{fig:posterior_sims5p}
\end{figure}

Figure \ref{fig:posterior_sims5p} displays the posterior distribution over allelic partitions for each prior and simulated scenario, compared to the true cluster size distributions, for a distortion probability value of $\psi_\ell = 0.05$ . Results for $\psi_\ell = 0.01$ are shown in Appendix \ref{app_sims1p}. In addition, Table \ref{tab:rates} displays the posterior average Jensen-Shannon (JS) distance between the MCMC samples of the partitions and the true partition, as well as more traditional RL classification error rates, i.e., False Negative Rate (FNR) and False Discovery Rate (FDR). The JS distance metric is based on a symmetrization of the Kullback–Leibler divergence, and allows us to evaluate how well the different models recover the true distribution of the allelic partition \citep{Lin91JensenShannon}. This is in contrast to the FNR and FDR values, which focus exclusively on pairwise comparisons. The JS distance values range between 0 and 1, so that values closer to zero are preferred.

\begin{table}[h!]
\setlength{\tabcolsep}{0.22em}
{\renewcommand{\arraystretch}{1.1}
{\fontsize{9}{10}\selectfont
\centering
\caption{Posterior average Jensen-Shannon (JS) distance, FNR and FDR (in percentages) for ESC and BBAP models for five scenarios simulated with distortion levels $\psi_\ell = 0.01$ and $0.05$.} \label{trates_simulations}
\begin{tabular}{c|c||ccc||ccc||ccc||ccc||ccc}
\multicolumn{1}{c}{}  & \multicolumn{1}{c}{} & \multicolumn{3}{c}{\bftab Scenario 1} & \multicolumn{3}{c}{\bftab Scenario 2}  & \multicolumn{3}{c}{\bftab Scenario 3}  & 
 \multicolumn{3}{c}{\bftab Scenario 4}  & \multicolumn{3}{c}{\bftab Scenario 5}  \\ 
  \hline
 & Prior & JS & FNR & FDR &JS & FNR & FDR & JS &FNR & FDR &JS & FNR & FDR &JS & FNR & FDR \\ 
  \hline
\parbox[t]{2.7mm}{\multirow{4}{*}{\rotatebox[origin=c]{90}{$\psi_\ell=0.01$}}} 
& ESCNB   &  0.026   & 2.6 & 1.1 & 0.037  & 3.6 & 3.5 & 0.043 & 2.9 & 0.1 & 0.056  & 4.4 & 1.4 & 0.063 &  5.5 &  2.1 \\ 
& ESCD      & 0.019   & 2.0 & 1.3 & 0.019  &  3.3 & 2.8 & 0.016 &1.4 & 0.1 & 0.013  &  1.8 & 0.8 & 0.024  &  3.3 &  0.8 \\ 
&BBAPG     &  0.023  & 2.3 & 1.2 & 0.037  & 3.3 &  3.7 & 0.018 & 1.5 & 0.1 & 0.029  & 2.2 & 1.1 &  0.041   & 4.0 &  1.4 \\ 
&BBAPNB   &  0.020  & 2.0 & 1.3& 0.039   & 2.9 &  4.2&  0.015 & 1.1 & 0.2 & 0.026   & 1.8 & 1.2 & 0.040  & 3.2 &  1.5 \\ 

  \hline
    \hline
\parbox[t]{2.7mm}{\multirow{4}{*}{\rotatebox[origin=c]{90}{$\psi_\ell=0.05$}}} 
& ESCNB   & 0.087 & 12.1 & 5.4 & 0.045 & 13.8 & 9.4 & 0.115 & 9.8 & 8.6 &  0.122  & 12.5 & 7.3 &  0.141  &  9.8 &  8.6 \\ 
& ESCD      & 0.050 & 9.5 & 4.6 & 0.040 &  13.2 & 9.5 & 0.053  & 7.8 & 8.5 &  0.041  &  8.1 &  5.9 &  0.080  &  7.8 &  8.5 \\ 
&BBAPG     & 0.068  & 10.7 & 4.8 & 0.045  & 12.9 &  9.7 & 0.062 & 8.1 & 8.1 & 0.073   & 8.9 & 6.8 &  0.101  & 8.1 &  8.1 \\ 
&BBAPNB   & 0.059    & 9.3 & 5.4 & 0.048 & 11.4 &  11.0 & 0.051 & 6.7 & 9.3 & 0.068 & 8.0 & 7.4 & 0.090  & 6.7 &   9.3 \\ 
 \hline
\end{tabular}
}
}
\end{table}

From Table \ref{tab:rates}, we observe that for the lowest level of distortion of the fields ($\psi_\ell=0.01$) all the models perform relatively well with FNR and FDR values between 0.1\% and  5.5\% in all cases. For $\psi_\ell=0.05$, where the RL task is expected to be more challenging, the error rates range between 4.6\% and 13.8\% for all simulated scenarios. Overall, ESCNB is the model with the worst performance across all three metrics, specially for simulated scenarios 3 to 5 which have less conventional cluster size distributions. Focusing on the results for $\psi_\ell=0.01$, where the clustering signal is stronger, we observe that ESCD seems to have the best performance in terms of the JS distance for all scenarios with very similar results from BBAPNB for scenarios 1 and 3. The performance in terms of FNR and FDR of the two BBAP calibrations is very similar to the ESCD for scenarios 1 and 3, while ESCD performs slightly better for the other scenarios. As the noise increases ($\psi_\ell=0.05$), the performance of ESCD and the two BBAP models becomes more similar. Note that all the models fail to capture the true maximum cluster size for all scenarios. In particular, the overall lower values of the JS distance for ESCD can be in part explained by its more accurate recovery of the true maximum cluster size compared to the other models (see Figure \ref{fig:posterior_sims5p}). Motivated by this and the natural data collection mechanisms of real applications, we explore the performance of a BBAP calibration that incorporates the maximum cluster size as prior information in section \ref{sec_illustrations}. Finally, when compared to each other, the results for BBAPG and BBAPNB display a trade-off between FNR and FDR. Although BBAPNB has a slight edge over BBAPG in terms of the JS distance, there is no clear outperforming model across all scenarios. This behavior highlights robustness of the results to different BBAP calibrations.

\section{Applications}\label{sec_illustrations}

In this section, we illustrate the behavior and performance of the BBAP and ESC models using the following three official statistics data sets.

\textbf{Durham}: The North Carolina State Board of Elections (NCSBE) provides snapshots of demographic information of
voters which are available to the public (\url{https://ncsbe.gov}). Using a snapshot from January of 2019, we consider a data set of 2,714 records of $K=2,000$ unique registered voters from Durham county. Duplicate records in this data commonly arise from individuals registering to vote after moving from a different county \cite{kaplan2018posterior, wortman2019record}. Ground truth about the partition is available through the NC Voter ID provided by the NCSBE. In order to perform record linkage we employ the following six fields of information: age,  sex, race, birth place, and first and last name initials.

\textbf{SDS}: The Social Diagnosis Survey (SDS) is a panel research project that studies indicators of quality of life in households in Poland (\url{http://www.diagnoza.com/index-en.html}). We consider a data set of $K=2,000$ unique individual members of households that participated in the survey in at least one of the years 2011, 2013, and 2015. Duplicate records occur longitudinally across the three waves but not within a specific year for a total of 3,574 records in the data. The data is available in horizontal format providing ground truth for the partition. We use six fields of information for RL: sex, date of birth (day, month and year), province of residence, and education level. 

\textbf{SIPP}: The Survey of Income and Program Participation (SIPP) is a longitudinal survey that collects information about the income and participation in federal, state, and local programs of individuals and households in the United States \citep{SIPP}. The data is publicly available through the Inter-university Consortium for Political and Social Research (ICPSR) (\url{https://www.icpsr.umich.edu}). We consider a data set of $K=1,000$ unique individuals interviewed over five waves of the survey performed between 2005 and 2006. The data contains a total of 4,116 records from individuals that are only duplicated across waves (not within). We use five fields of information for RL: sex, year and month of birth, race, and state of residence.

\begin{figure}[h!]
\includegraphics[width=\textwidth]{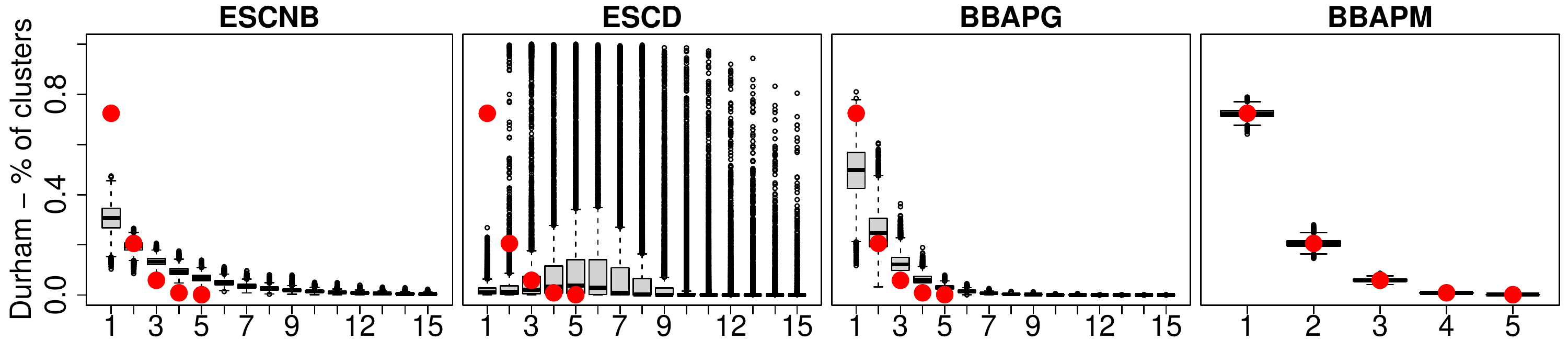}
\includegraphics[width=\textwidth]{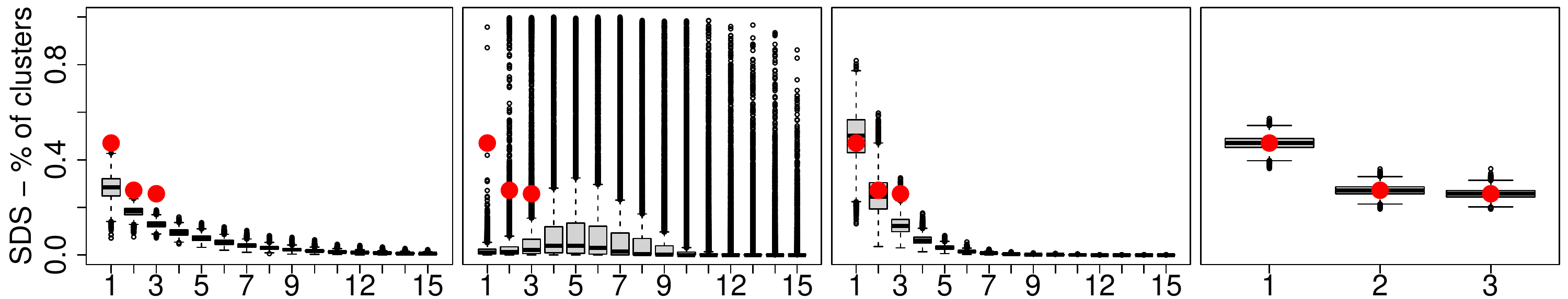}
\includegraphics[width=\textwidth]{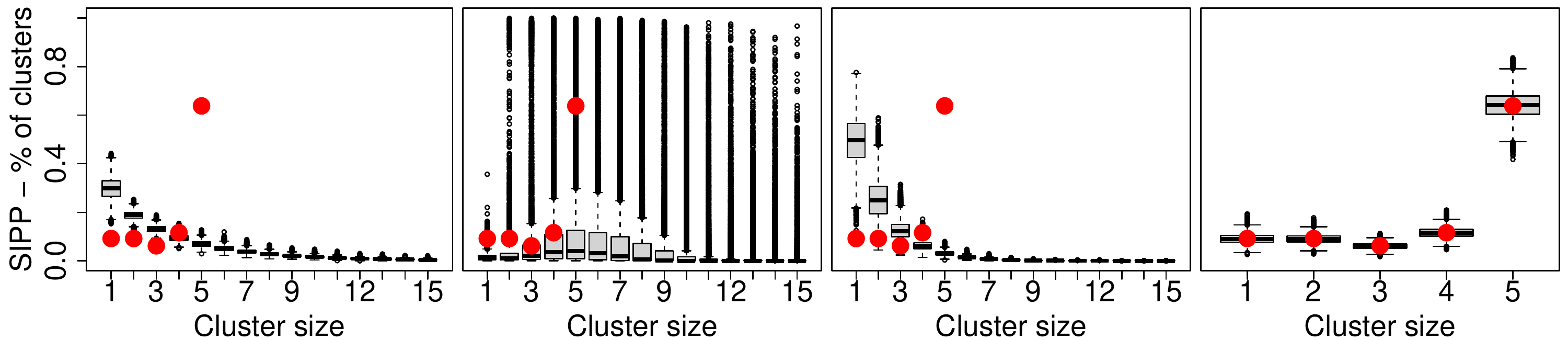}
\caption{Prior distribution of the allelic partition (boxplots) and true data partition (red dots) for ESC and BBAP models for the Durham, SDS and SIPP data sets.} \label{fig:prior_samples}
\end{figure}

In contrast to the Durham data, the SDS and SIPP datasets intrinsically provide prior information about the expected maximum cluster size in the partition due to their panel structure. Indeed, given the number of waves in each survey we expect the size of the largest clusters to be three and five for SDS and SIPP, respectively. Although these illustrations do not necessarily reflect the conditions of real data applications where ground truth might not be available, we use these data sets to display the adaptability of the BBAP model to prior knowledge at different scales. For this purpose, we consider two different calibrations of the BBAP model for all datasets. First, the default Geometric specification with $\piv=\Geom{(0.5)}$ using $M^*=15$ -- denoted as BBAPG. Second, an informed specification where $\piv$ reflects the true data partition and $M^*$ is fixed at the true maximum cluster size -- denoted as BBAPM. To perform the elicitation of the hyperparameters, we use coefficients of variation of $25\%$ and $5\%$ for BBAPG and BBAPM, respectively.

For the ESC models,  similar to the simulation studies, we set $\alpha = 1$,  $a \sim \text{Gamma}\left(1, 1\right)$, and  $q \sim \text{Beta}\left(2, 2\right)$ and work with a truncated version of the ESC models with $M^*=100$. Finally, we assume a Beta prior distribution with mean 0.01 and standard deviation of 0.01 for the distortion probabilities of the fields, $\psi_\ell$, for all the models (see  Section \ref{sec_rlmodel}). Figure \ref{fig:prior_samples} displays samples from all the prior distributions against the true allelic partition for each dataset (ESC results are shown up to $M^*=15$ for visibility). Durham data displays the more traditional geometric-like behavior of the true allelic partition, while the SDS and SIPP partitions are less conventional. Evidently, the prior belief for the SIPP data is extremely misspecified under all the non-informed prior models i.e. excluding the BBAPM calibration. Consistent with what we observed in section \ref{sec_sims},  ESCNB and BBAPG behave similarly (BBAPG has a faster rate of decay), while the ESCD prior behavior is quite different and in this case favors configurations in which the most frequent cluster size is between 5 and 6. On the other hand, the BBAPM calibration is designed to match the true allelic partition quite closely. 

All results presented below are based on 20,000 samples from the combination of two chains of 10,000 iterations, obtained after a burn-in period of 10,000 samples for each chain. Traceplots used for convergence diagnostics for the BBAPG model are included in Appendix \ref{app_converge}.

\subsection{Results} \label{sec_results}

\begin{figure}[h!]
\includegraphics[scale=0.5]{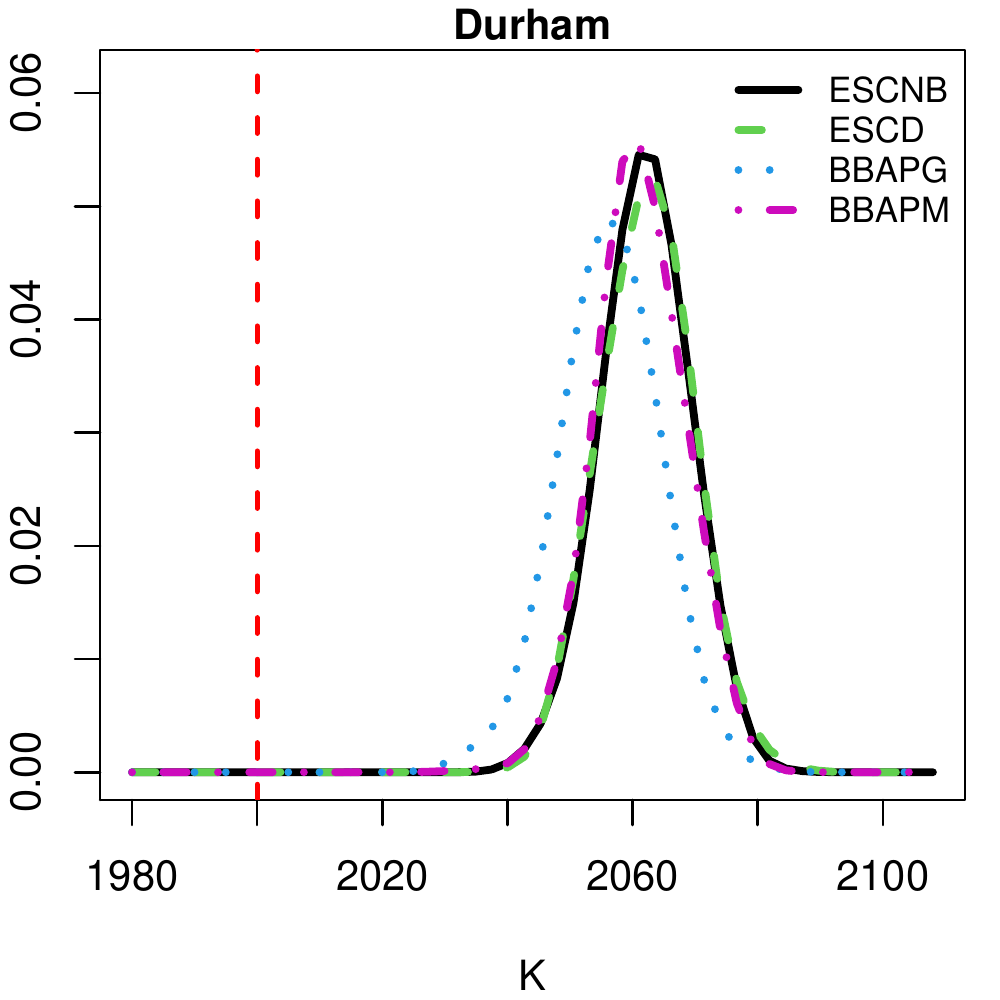} 
\includegraphics[scale=0.5]{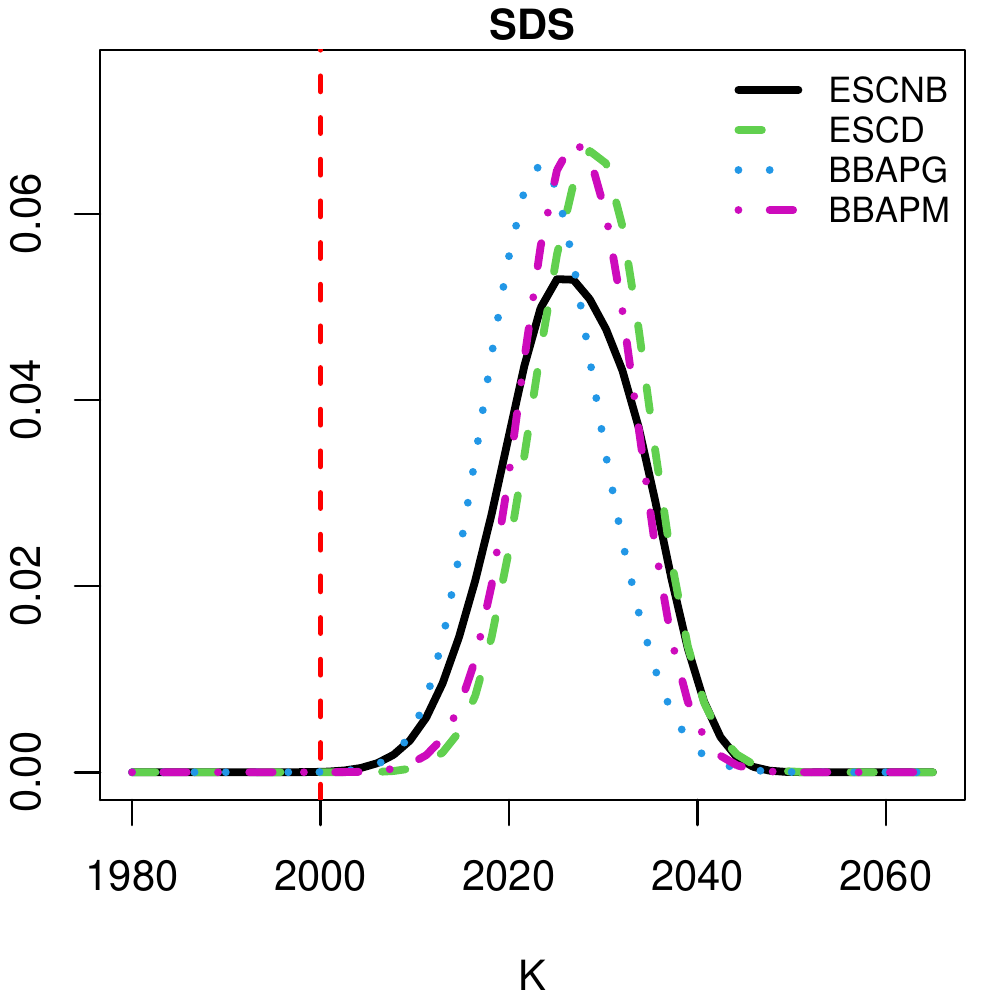} 
\includegraphics[scale=0.5]{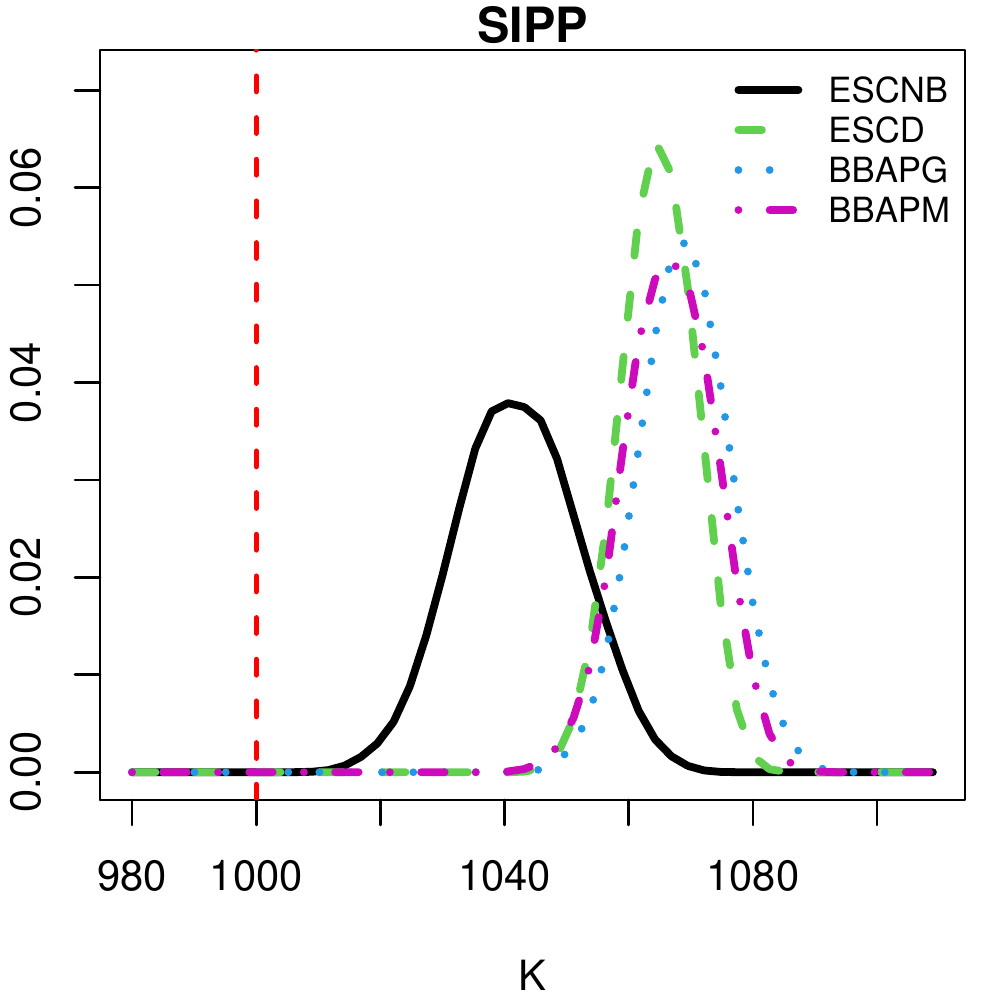}
\caption{Posterior distribution of the number of clusters ($K$) for ESC and BBAP models for the Durham, SDS and SIPP data sets. The vertical line represents the true number of clusters in each application.} \label{fig:AllK}
\end{figure}

\begin{figure}[h!]
\includegraphics[width=\textwidth]{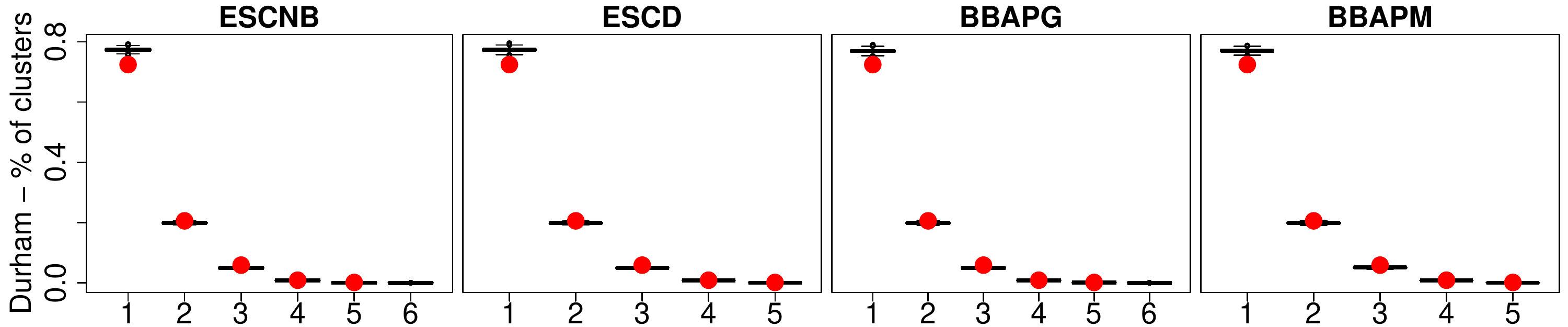}
\includegraphics[width=\textwidth]{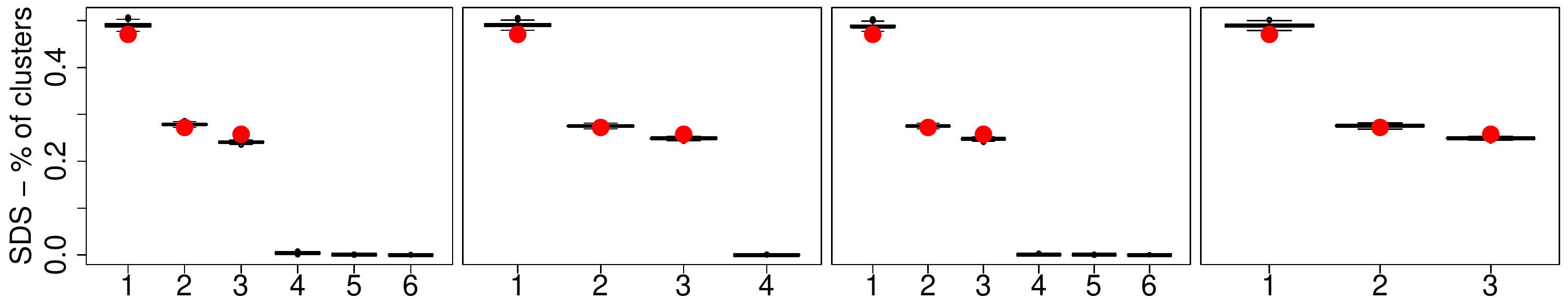}
\includegraphics[width=\textwidth]{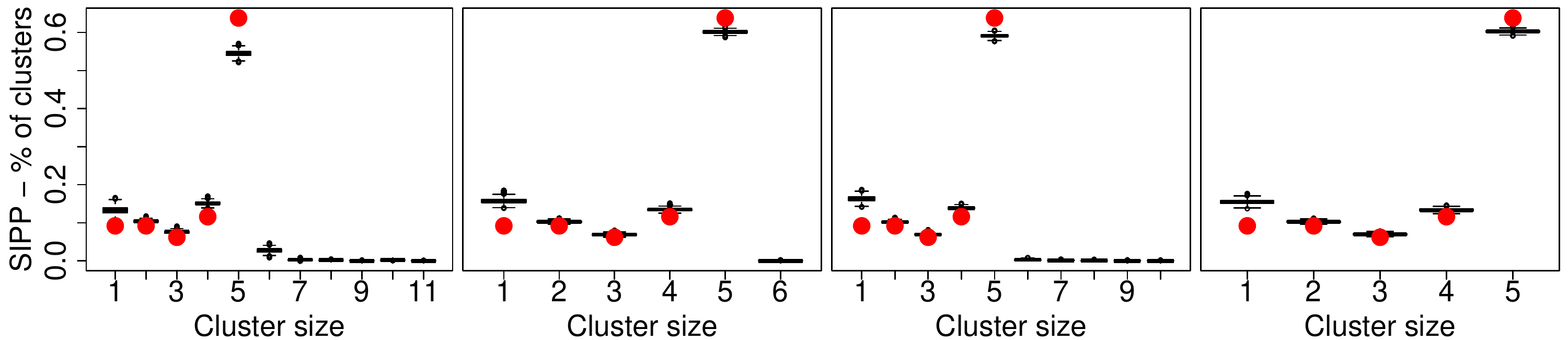}
\caption{Posterior distribution of the allelic partition (boxplots) and true data partition (red dots) for ESC and BBAP models for the Durham, SDS and SIPP data sets.} \label{fig:Allsizes}
\end{figure}

Figure \ref{fig:AllK} shows the posterior distribution of the number of clusters (i.e., the number of unique individuals in the dataset) under each prior and dataset. Note that the models fail to capture the true number of clusters by consistently overestimating it in all cases.  However, BBAPG seems to have a slightly more accurate performance in the Durham and SDS datasets.  Interestingly, it is the ESCNB prior that provides the most accurate estimate of the number of unique individuals for the SIPP dataset. This seems to be due to an overestimation in the number of clusters of size 5 (see Figure \ref{fig:Allsizes} and the explanation below). 

Figure \ref{fig:Allsizes} displays the posterior distribution over allelic partitions for each prior and data set, and compares them against the truth. In addition, Table \ref{tab:rates} displays the posterior average JS distance, FNR and FDR. From Table \ref{tab:rates}, we observe that the FNR values for the Durham data are the highest for all the datasets (above 13\%), compared to values below 5.2\% for the SDS and SIPP applications. On the other hand, the FDR values are below 4.4\% for all models and data sets. The largest JS distances are observed for the SIPP dataset, while the lowest ones are seen in SDS.

\begin{table}[h!]
\centering
\caption{Posterior average Jensen-Shannon (JS) distance, FNR and FDR (in percentages) for ESC and BBAP models for the Durham, SDS and SIPP data sets.} \label{tab:rates}
{\small
\setlength{\tabcolsep}{0.3em}
\renewcommand{\arraystretch}{1}
\begin{tabular}{c||ccc||ccc||ccc}
\multicolumn{1}{c}{}    & \multicolumn{3}{c}{\bftab Durham} & \multicolumn{3}{c}{\bftab SDS}  & \multicolumn{3}{c}{\bftab SIPP} \\
\hline
Prior & JS & FNR & FDR & JS &  FNR & FDR & JS & FNR & FDR \\
\hline 
ESCNB  & 0.025 &  13.7 & 3.5 & 0.042 & 4.1 & 2.7 & 0.129 & 5.2 & 4.4 \\ 
ESCD & 0.028 &  13.9 & 3.4 &  0.011 &  3.8 & 1.7 & 
0.067 & 4.8 & 1.8 \\
BBAPG &  0.023 &  13.0 &  4.1 & 0.025 & 3.7 & 2.2 & 
0.084 & 4.9 & 2.1 \\
BBAPM & 0.024 & 13.3 & 3.8 & 0.011 &  3.8 & 1.7 & 
0.066 & 4.6 & 1.7 \\
\hline 
\end{tabular}}
\end{table}

All priors perform similarly for the Durham dataset, which has the more traditional allelic partition distribution. In spite of the very similar performance, BBAPG seems to have a slight edge over ESCNB and ESCD in terms of the mean JS distance and FNR, at the price of a slightly higher FDR. The reason seems to be that BBAPG is more aggressive in terms of encouraging the creation of non-singleton clusters (see  Figure \ref{fig:prior_samples}). Note that this result is consistent with our previous observation that BBAPG seems to perform slightly better in terms of estimating the number of unique individuals in the sample for this dataset. BBAPM (the ``informed'' prior) has a very similar performance to BBAPG in the Durham dataset. On the other hand, in the SDS and SIPP datasets, ESCNB tends to underperform across all three metrics. Among the other two ``uninformed'' models, ESCD seems to have the best performance in terms of the JS distance and FDR, but the behavior in terms of the FNR is very similar to that of BBAPG. Finally, the behavior of BBAPM is very similar to that ESCD in these two datasets, although BBAPM seems to exhibit a slightly better FNR and FDR than ESCD for the SIPP dataset.  Consistently with the simulation study,  we find that overall our proposed prior performs better than ESCNB and displays competitive results compared to ESCD.

\subsection{Point estimation} \label{sec_point}

As in other clustering applications, finding a unique optimal partition of the data is of interest for RL problems. In many cases, in addition to estimation of the number of unique entities, RL is also a required preprocessing step for subsequent statistical analysis with the linked data  \citep{gutman-2013, sadinle-2014, sadinle2018, HofRavelliZwinderman17, kaplan2018posterior}. In the microlustering context, summarizing the information provided by a sample of partitions into an optimal partition is specially challenging. The large number of small clusters expected in the posterior samples of the partition and the high-dimensionality of the space in real data scenarios compels the use of scalable algorithms to find the optimal partition. Decision theoretical approaches for optimal Bayesian estimation that rely on loss functions functions for the space of partitions include those of \cite{lau-2007}, \cite{wade_cluster_2018} and \cite{Rastelli2018}. Here, we utilize the approach of  \cite{Rastelli2018} who proposes a scalable greedy algorithm to minimize
the expected posterior loss (EPL) under different loss functions. Table \ref{tab:EPL} displays the estimated number of clusters (K), the JS distance and the error classification rates for the point estimates of the partition obtained with the greedy EPL algorithm for Binder's (B), the Normalised Information Distance (NID), and the Variation of Information (VI) loss functions. Results are based on the last 2,000 iterations of the posterior samples using the \texttt{R} package \texttt{GreedyEPL} \citep{GreedyEPL}.

\begin{table}[h!]
\centering
\caption{Estimated number of clusters (K), JS distance, FNR and FDR for point estimates of the partitions obtained with a greedy EPL algorithm for Binder's (B),  Normalised Information Distance (NID), and Variation of Information (VI) loss functions for Durham, SDS and SIPP. True number of clusters is K=2,000 for Durham and SDS, and K=1,000 for SIPP.} \label{tab:EPL}
{\small
\setlength{\tabcolsep}{0.31em}
\renewcommand{\arraystretch}{1}
\begin{tabular}{c|c||cccc||cccc||cccc}
\multicolumn{1}{c}{} & \multicolumn{1}{c}{}    & \multicolumn{4}{c}{\bftab Durham} & \multicolumn{4}{c}{\bftab SDS}  & \multicolumn{4}{c}{\bftab SIPP} \\
\hline
& Prior & K & JS & FNR & FDR & K & JS &  FNR & FDR & K & JS & FNR & FDR \\
\hline 
 \parbox[t]{7mm}{\multirow{4}{*}{\rotatebox[origin=c]{0}{B}}} 
& ESCNB  & 2063 & 0.025 &  12.4 & 1.9 & 2024 & 0.042 & 4.0 & 2.7 & 
1053 & 0.104 & 4.3 & 2.8 \\ 
& ESCD & 2060 & 0.024 &  12.1 & 2.0 & 2026 &  0.010 &  3.4 & 1.3 & 
1066 & 0.068 & 4.6 & 1.6 \\
& BBAPG & 2057 &  0.023 &  11.5 &  2.2 & 2019 & 0.025 & 3.0 & 1.8 & 
1072 & 0.083 & 4.7 & 1.6 \\
& BBAPM & 2062 & 0.025 & 11.7 & 1.5 & 2028 & 0.011 &  3.4 & 1.2 & 
1070 & 0.076 & 4.4 & 1.3 \\
\hline \hline 
\parbox[t]{7mm}{\multirow{4}{*}{\rotatebox[origin=c]{0}{NID}}} 
& ESCNB  & 2063 & 0.028 &  12.7 & 3.1 & 2027 & 0.045 & 4.2 & 3.1 & 
1064 & 0.137 & 5.0 & 6.2 \\ 
& ESCD & 2062 & 0.025 &  12.2 & 1.7 &  2033 & 0.027 &  3.7 & 1.9 & 
1070 & 0.089 & 4.7 & 2.6 \\
& BBAPG & 2057 &  0.025 &  11.7 &  3.3 & 2023 & 0.034 & 3.4 & 2.6 & 
1081 & 0.109 & 4.6 & 2.9 \\
& BBAPM & 2062 & 0.026 & 11.8 & 2.0 & 2031 & 0.032 &  3.8 & 2.0 & 
1077 & 0.100 & 4.7 & 2.8 \\
\hline \hline 
\parbox[t]{7mm}{\multirow{4}{*}{\rotatebox[origin=c]{0}{VI}}} 
& ESCNB  & 2031 & 0.057 &  12.4 & 16.0 & 1908 & 0.159 & 4.0 & 25.2 & 
971 & 0.200 & 4.4 & 20.4 \\ 
& ESCD & 2023 & 0.062 &  12.1 & 18.7 & 1912 &  0.154 &  3.4 & 24.4 & 
982 & 0.191 & 4.3 & 19.3 \\
& BBAPG & 2016 &  0.067 &  11.7 &  20.7 & 1892 & 0.163 & 3.0 & 26 & 
982 & 0.200 & 4.2 & 20.2 \\
& BBAPM & 2030 & 0.056 & 11.7 & 15.6 & 1932 & 0.139 &  3.5 & 21.6 & 
975 & 0.194 & 4.1 & 20.0 \\
\hline 
\end{tabular}}
\end{table}

Based on the results for all three data sets and prior distributions, we find that using the VI loss in the microclustering context of RL is not advisable. The greedy EPL with VI loss consistently underestimates the number of clusters compared to B and NID. Even though this behavior leads VI to an estimated number of clusters that is closer to the truth for Durham and SIPP,  the overclustering also results in overinflated JS distance and FDR values (above 15\% for all datasets and priors). As discussed in \cite{Rastelli2018}, the greedy EPL with VI loss is the most suitable for conventional clustering applications where the number of clusters is not too large. For microclustering, however,  we observe that greedy EPL with Binder's loss yields the best performance in terms of lower JS distance and error rates as it tends to slightly overestimate the number of clusters compared to VI. The performance of greedy EPL with NID comes as a close second to Binder's loss for all data sets and priors.

\section{Discussion}\label{sec_discussion}

We have developed a new prior specification for the linkage structure in record linkage problems based on allelic partitions. Our approach is computationally tractable and permits easy incorporation of prior information. Our experiments show that our formulation performs competitively compared to the existing state-of-the-art microclustering models when prior information is not available, and can outperform state-of-the-art alternatives when accurate prior information is available. We have also introduced a set of novel microclustering conditions, which provides a unified framework for thinking about prior specification in applications such as RL where the number of clusters is expected to grow linearly with the number of observations. 

Our work opens up several doors for future research. Scalability is still the main challenging aspect of big data applications of RL involving Bayesian models. Real world data sets, such as the NCSBE voter registration data discussed in Section \ref{sec_illustrations}, can contain millions of records leading to a high-dimensional space of partitions. A crucial aspect of future work involves the development of computational algorithms for efficient posterior inference in the microclustering setting using, for example, Metropolis-Hastings (MH) schemes with better properties \citep{zanella2019informed} or fast computation techniques in the domain of variational approaches \citep{broderick-2014, Blei2017variational}. The computational limitations also extend to the implementation of scalable algorithms for optimal Bayesian estimation of the partitions in such high-dimensional discrete spaces.   

\bibliographystyle{apalike}
\bibliography{references}

\clearpage

\appendix

\section{Additional simulation results }\label{app_sims1p}

The following plots display the posterior distributions of the allelic partitions for the five simulated scenarios generated with a distortion probability value of 1\% for the fields (see Section \ref{sec_sims}). 

\begin{figure}[h!]
\includegraphics[width=\textwidth]{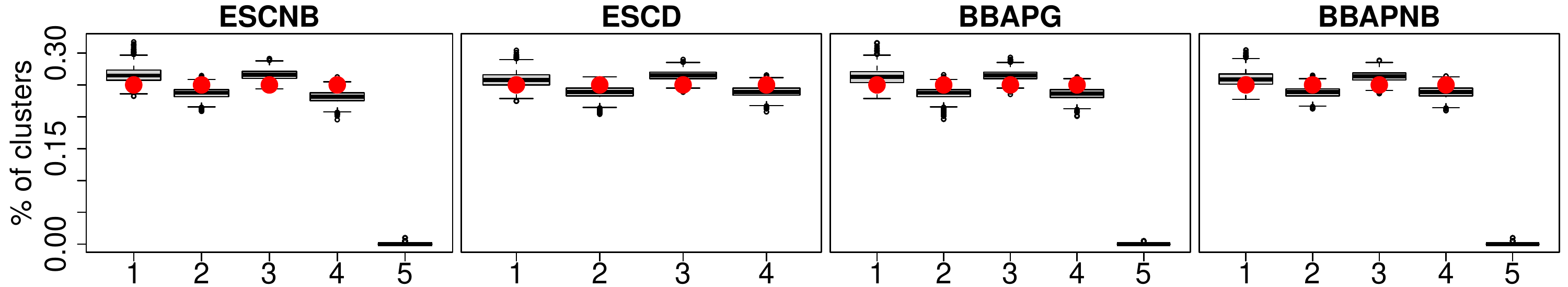}
\includegraphics[width=\textwidth]{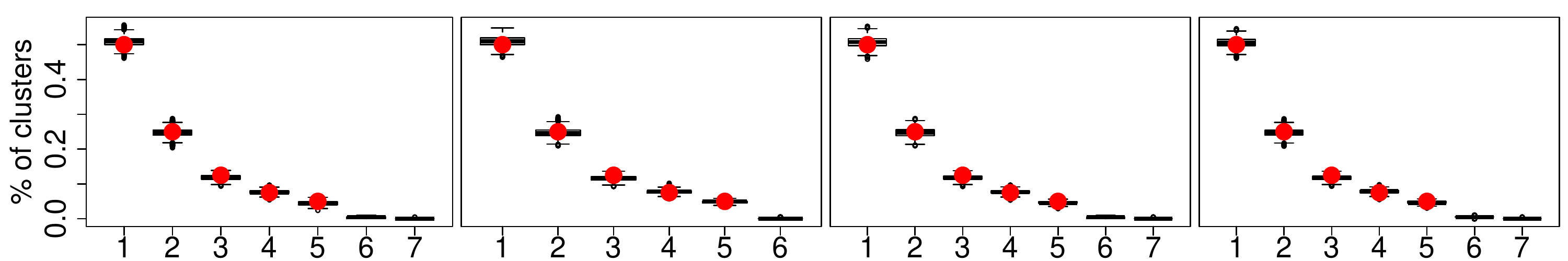}
\includegraphics[width=\textwidth]{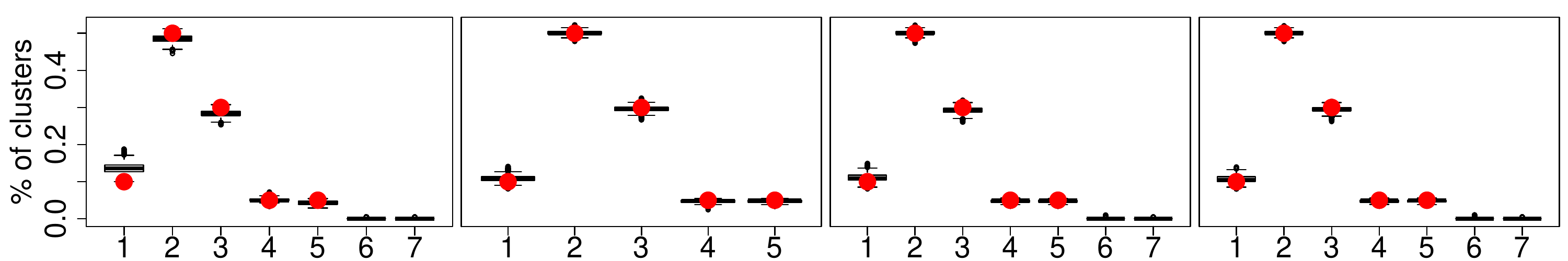}
\includegraphics[width=\textwidth]{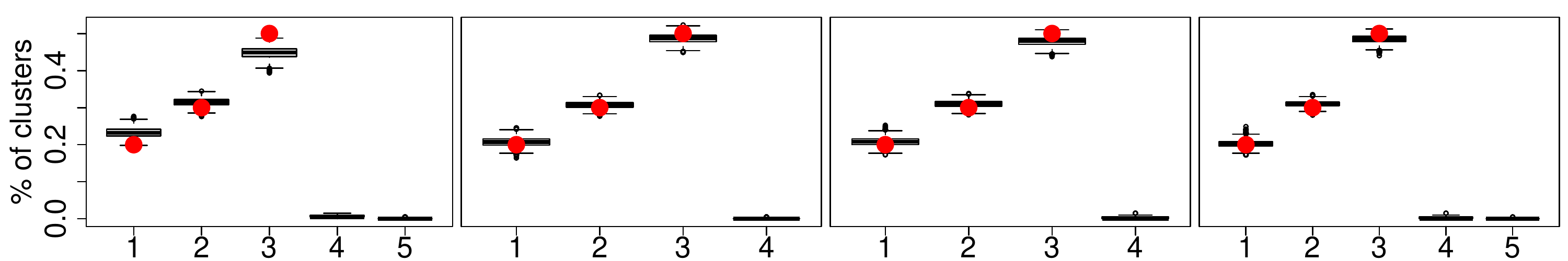}
\includegraphics[width=\textwidth]{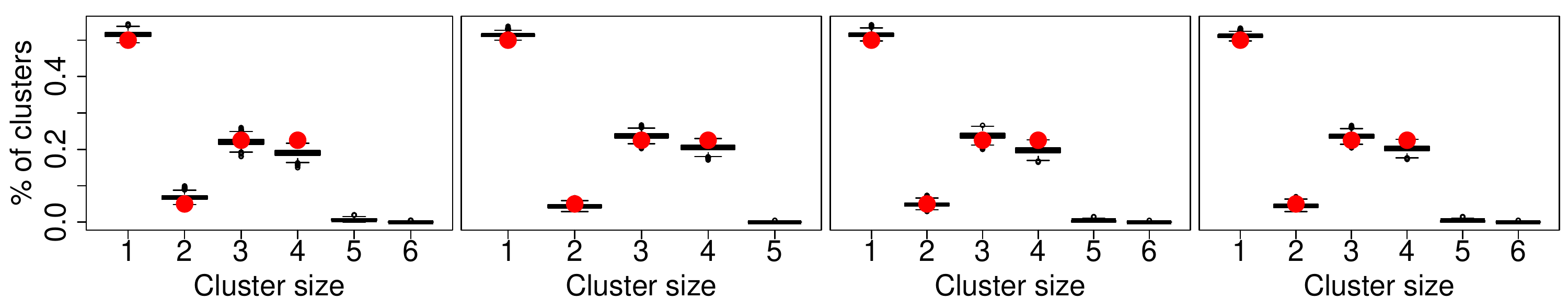}
\caption{Posterior distribution of the allelic partition (boxplots) for ESC and BBAP models, and true data partition (red dots) for five simulated scenarios with distortion probability $\psi_\ell=0.01$.} \label{fig:posterior_sims1p}
\end{figure}

\section{Convergence diagnostics}\label{app_converge}

\begin{figure}[h] 
\includegraphics[width=0.32\textwidth]{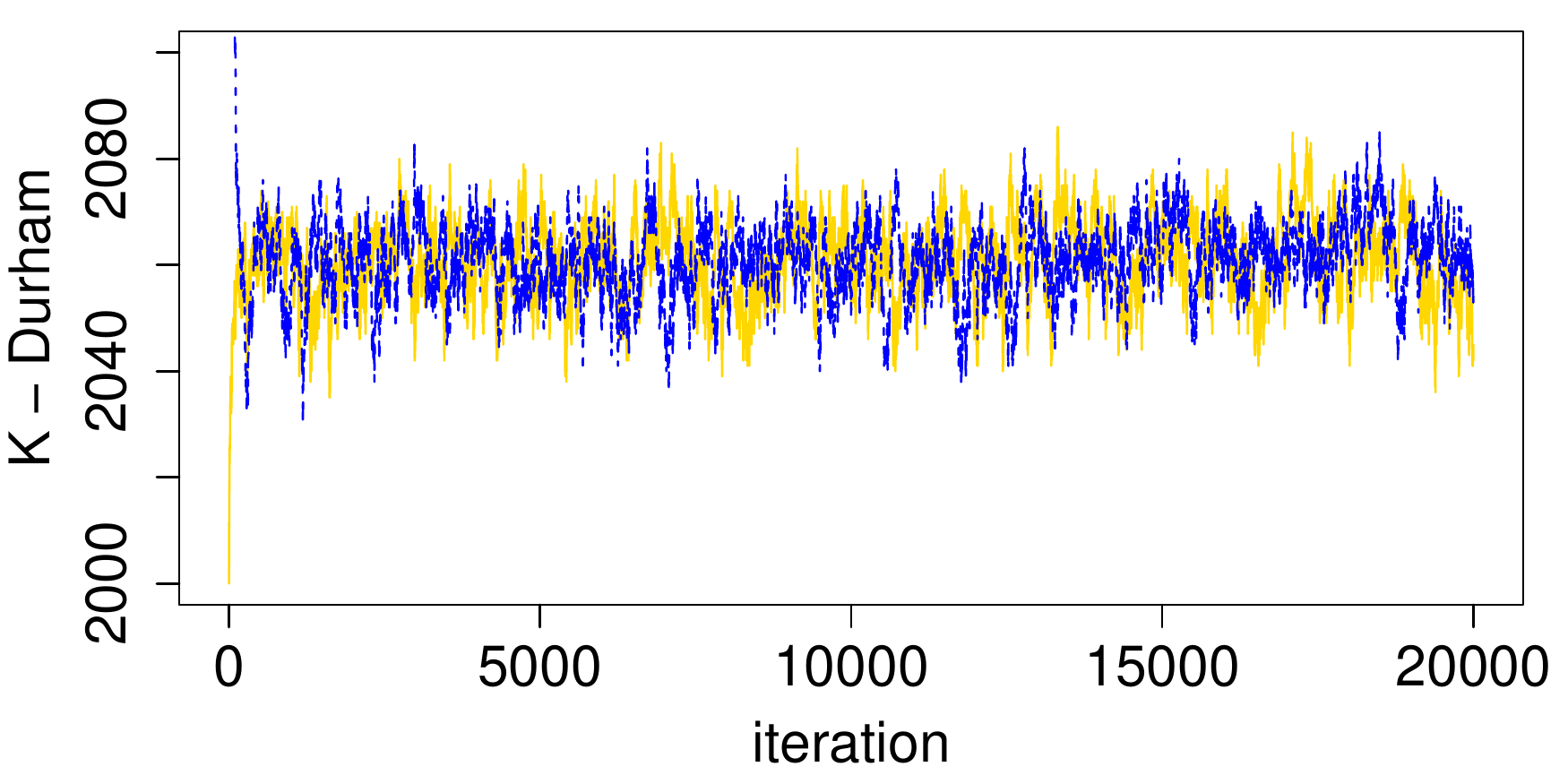}
\includegraphics[width=0.32\textwidth]{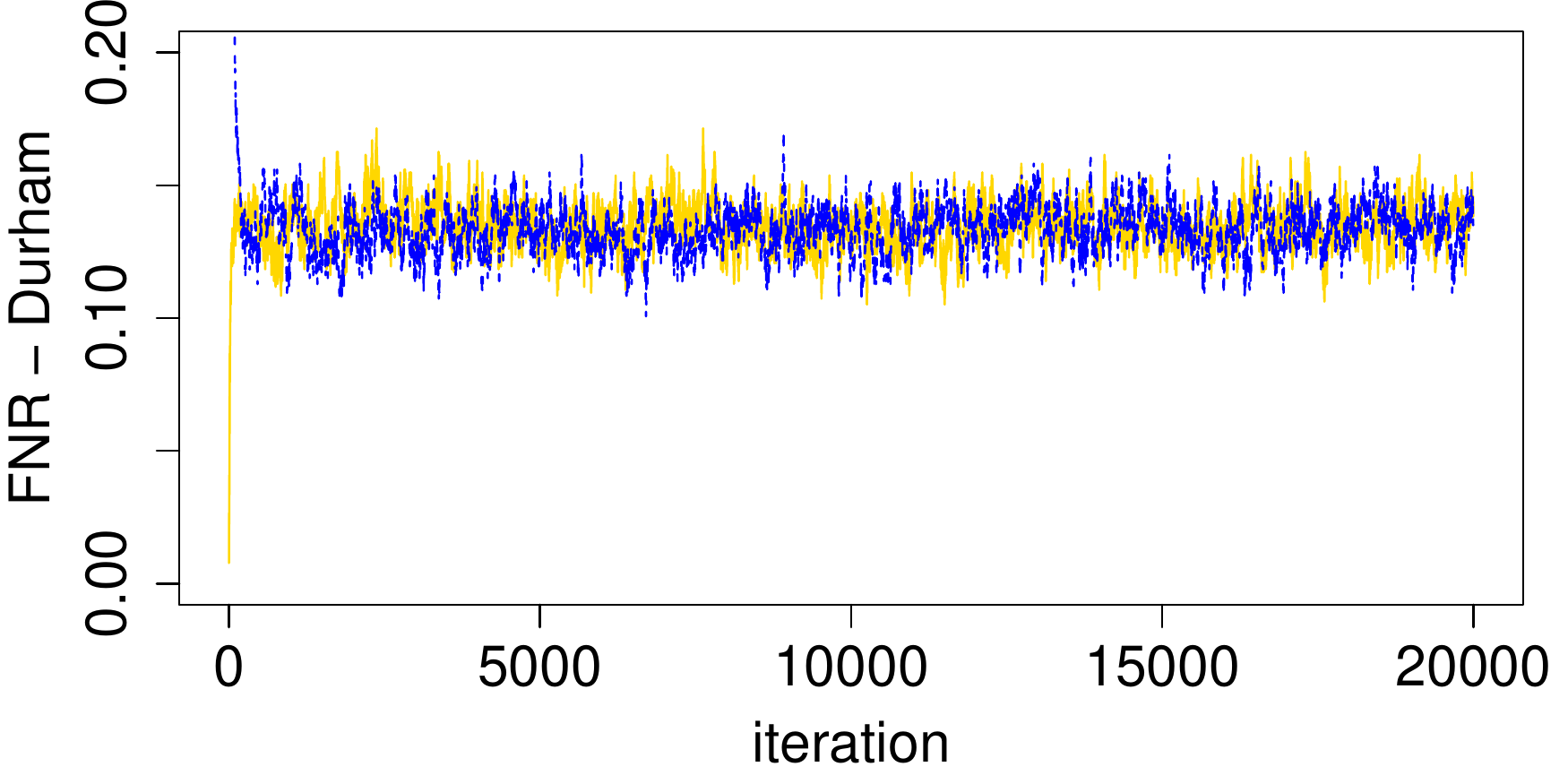}
\includegraphics[width=0.32\textwidth]{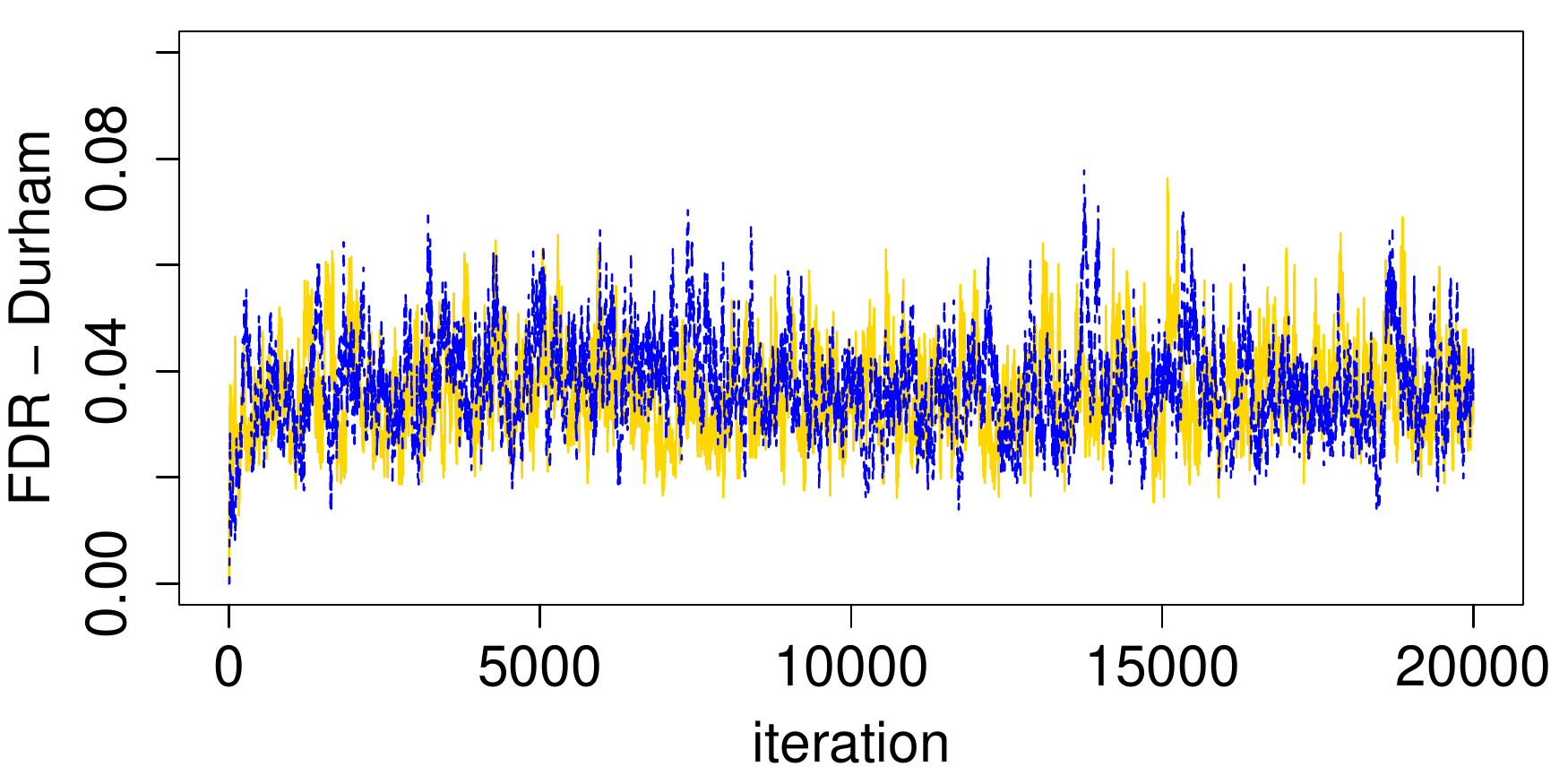}
\includegraphics[width=0.32\textwidth]{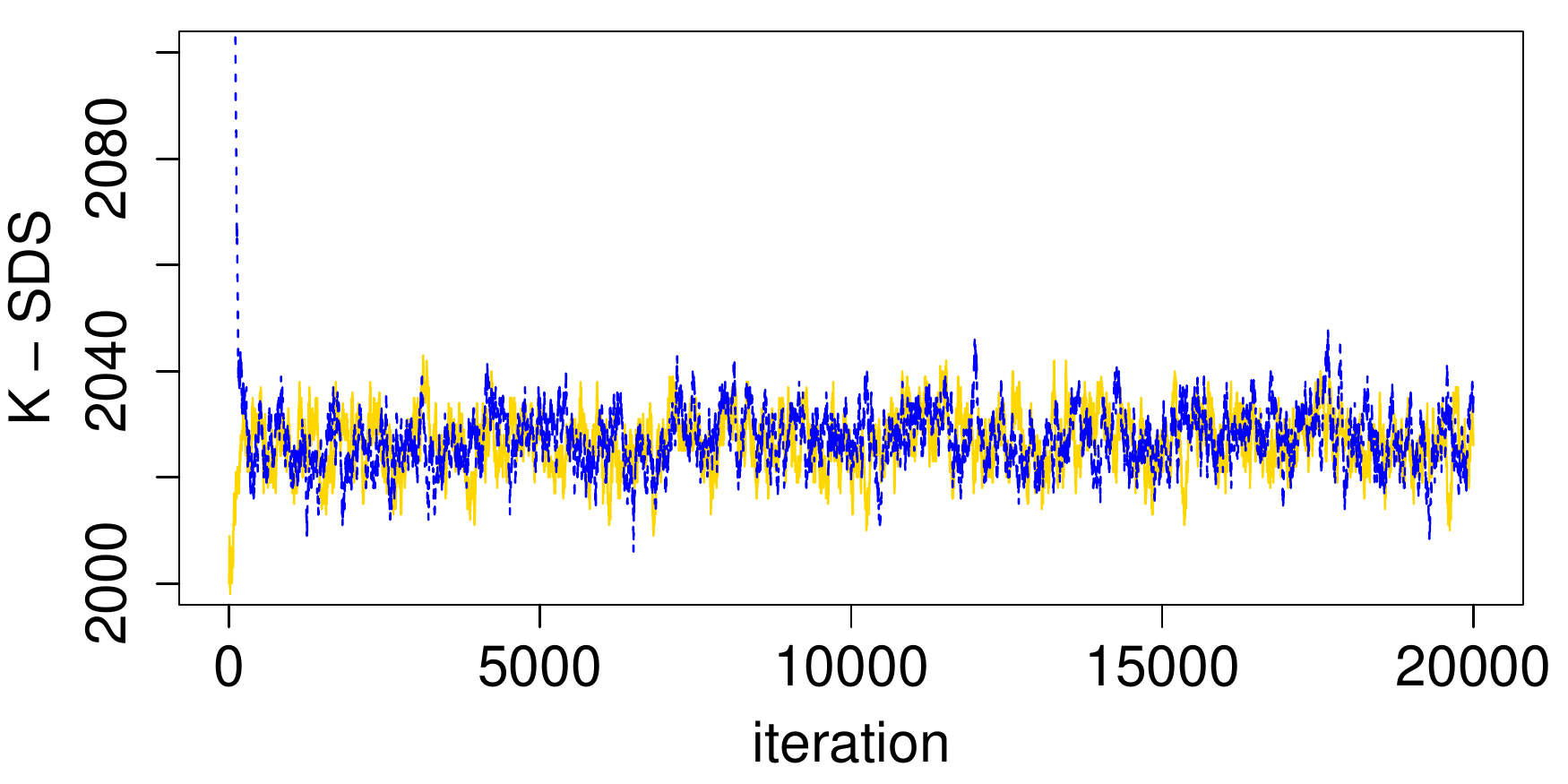}
\includegraphics[width=0.32\textwidth]{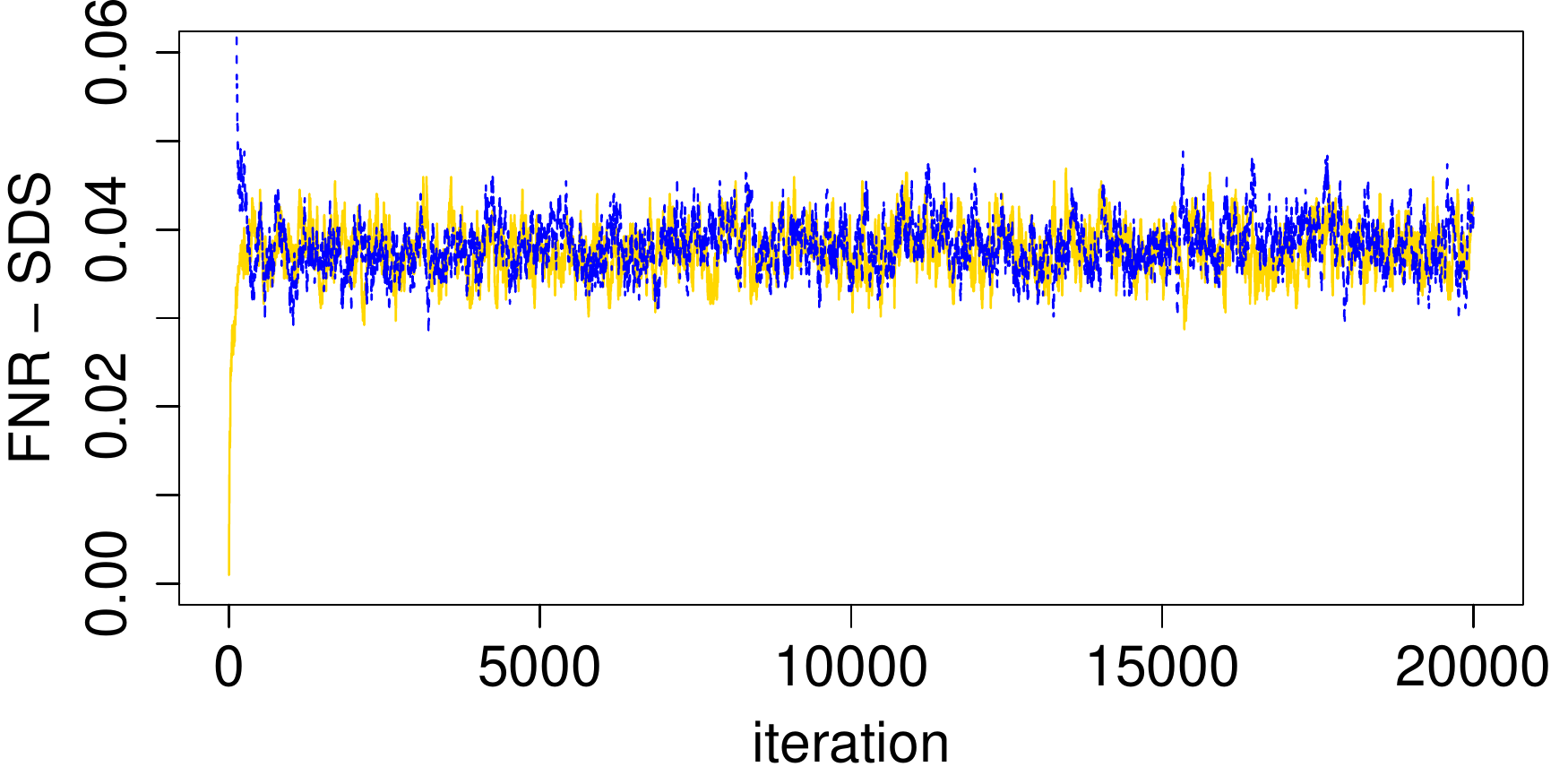}
\includegraphics[width=0.32\textwidth]{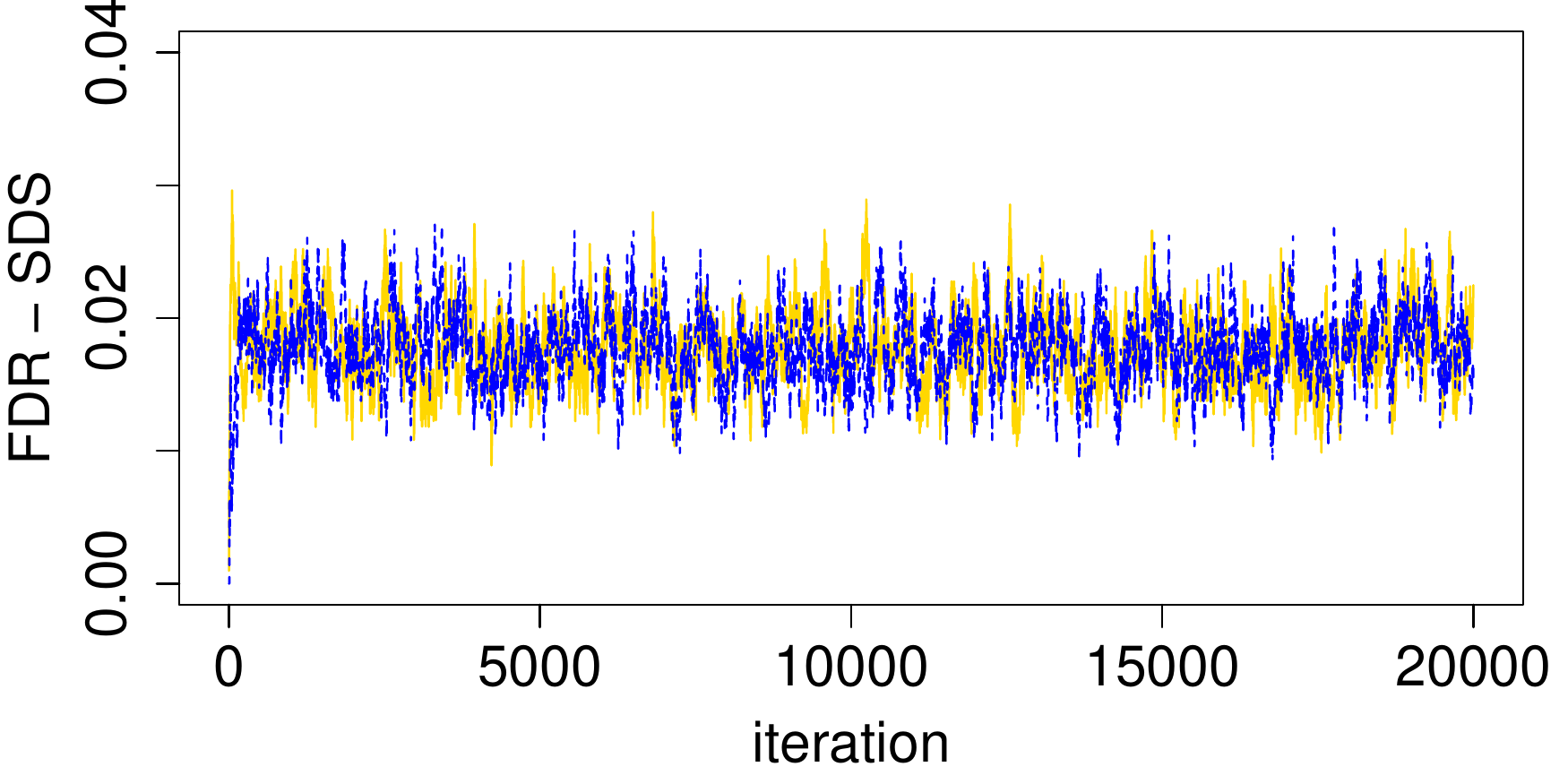}
\includegraphics[width=0.32\textwidth]{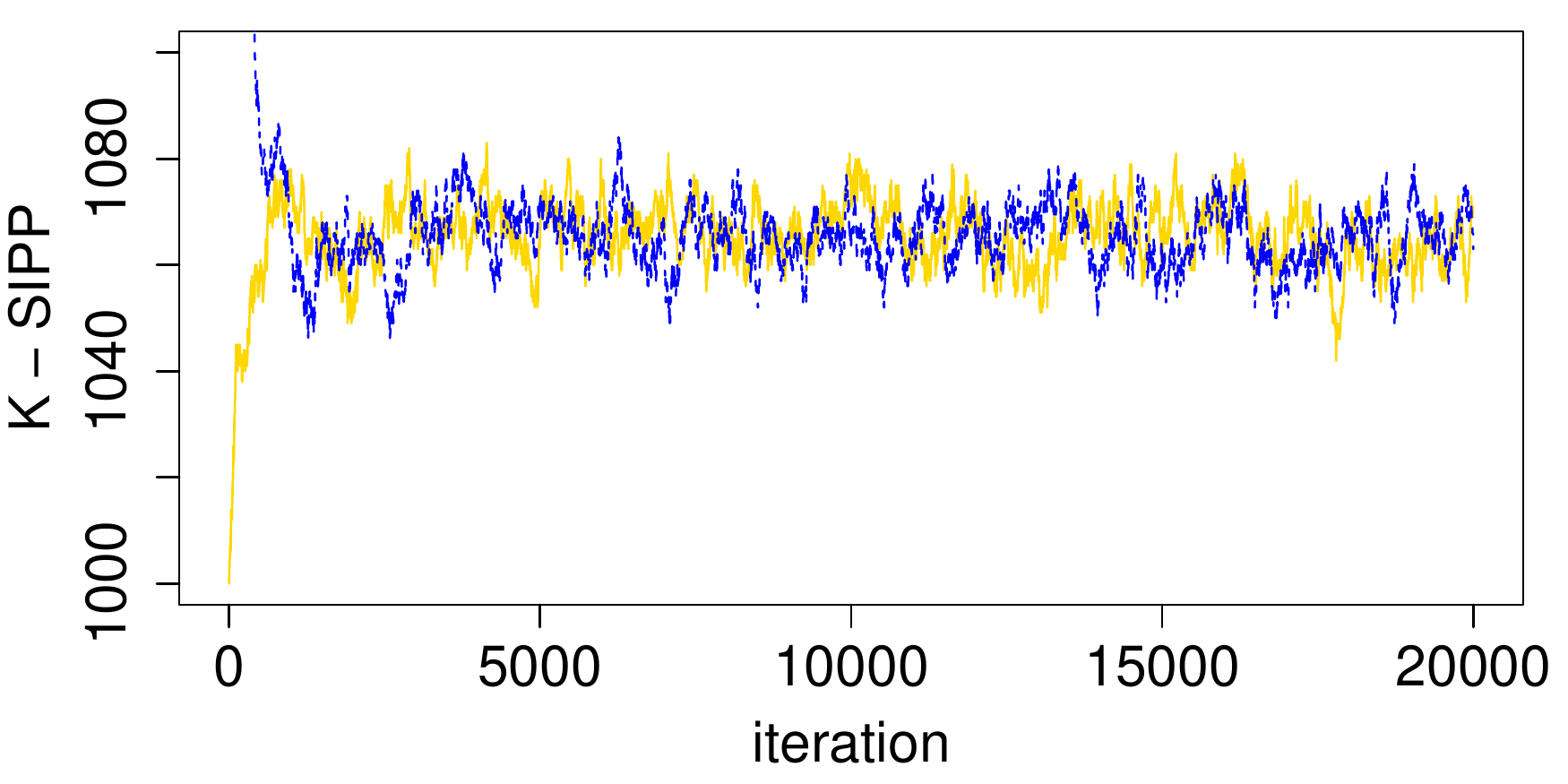}
\hspace{1mm}
\includegraphics[width=0.32\textwidth]{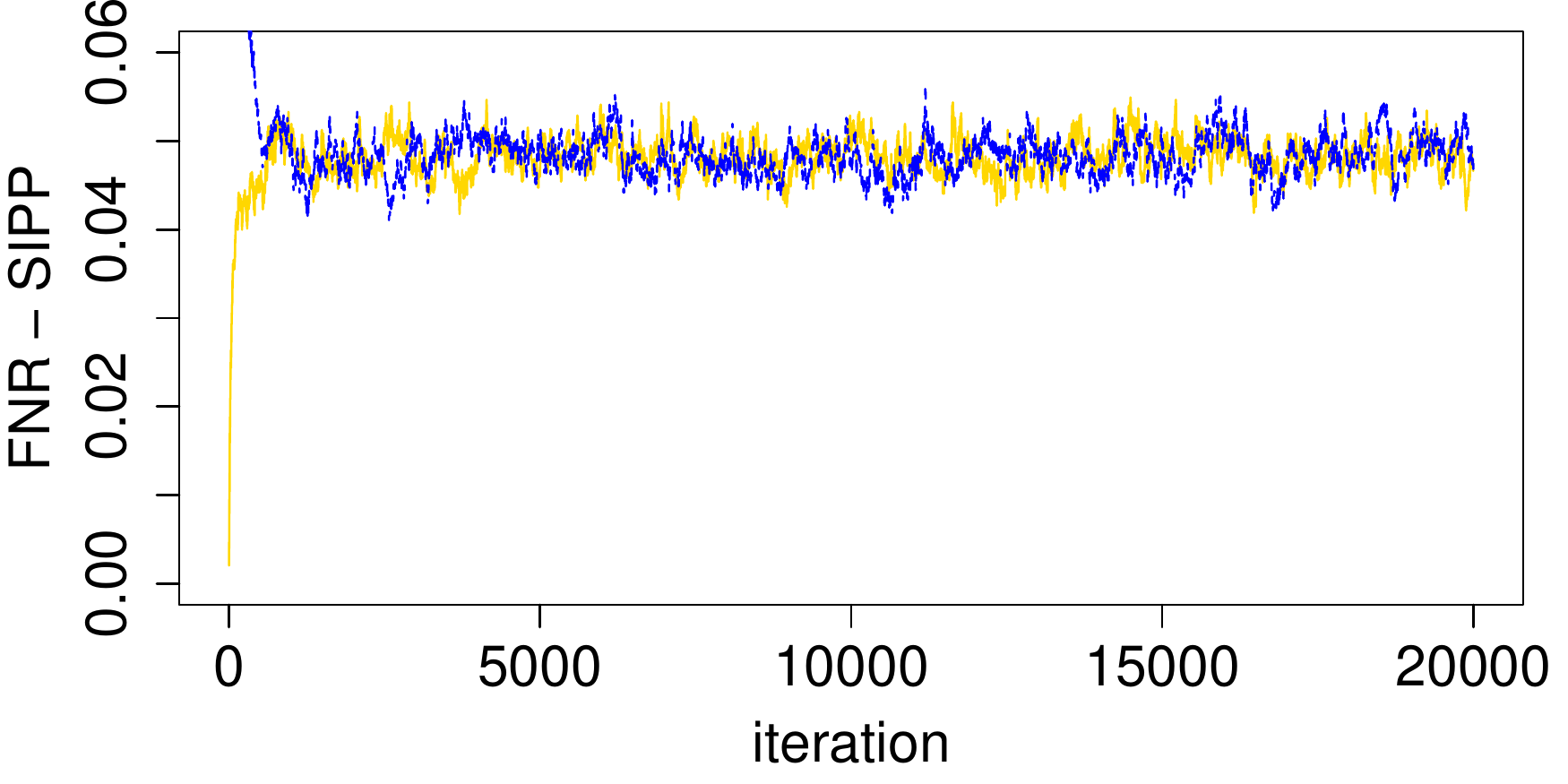}
\hspace{1mm}
\includegraphics[width=0.32\textwidth]{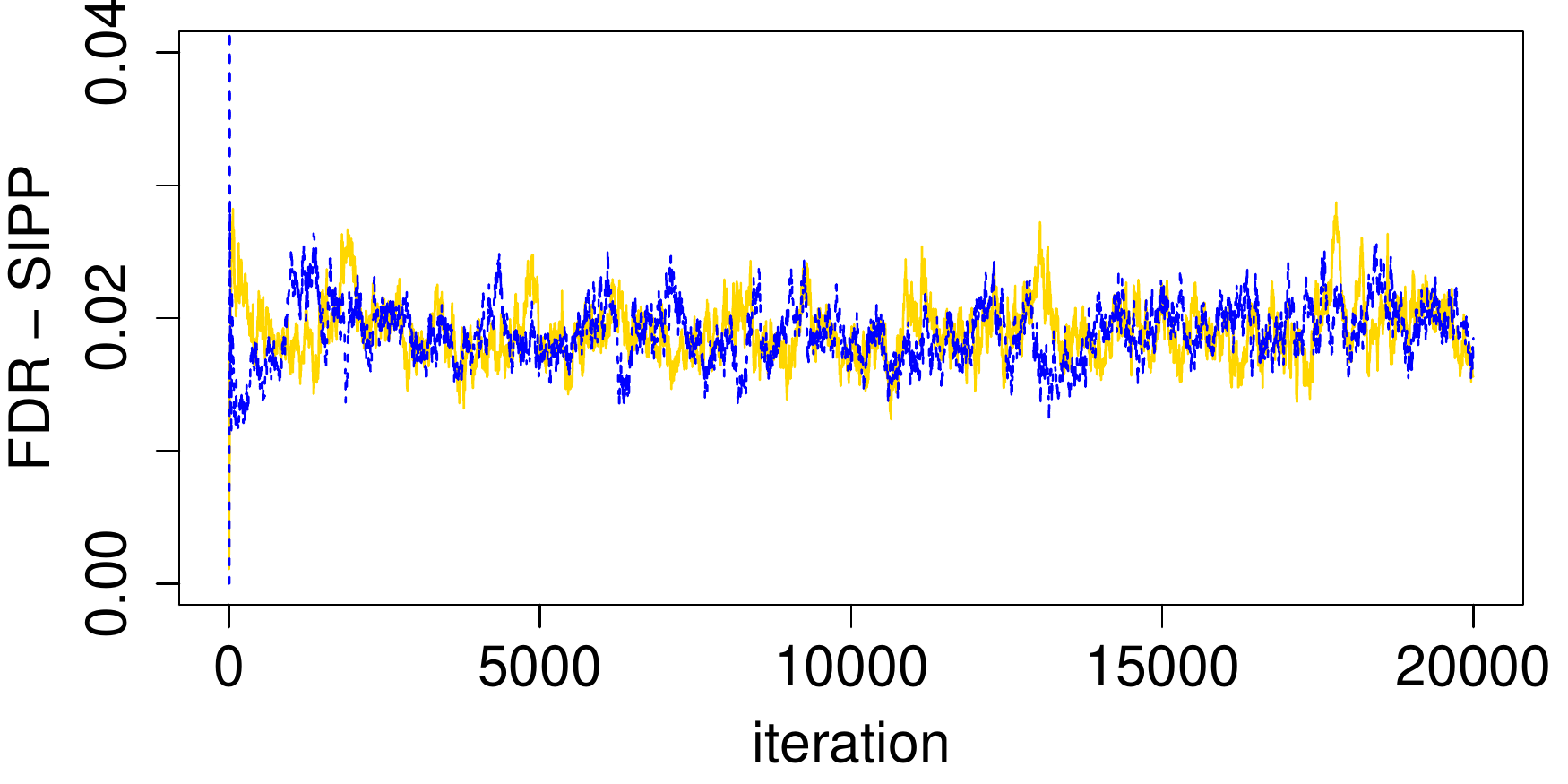}
\caption{Trace plots of number of clusters (K), false negative rate (FNR) and false discovery rate (FDR) for two chains of 20,000 iterations of the BBAPG model for Durham, SDS and SIPP data sets, respectively.} \label{fig:Allconverge}
\end{figure}

Figures \ref{fig:Allconverge} displays the traceplots for K, FNR and FDR for two chains of the BBAPG model for the Durham, SDS and SIPP data sets discussed in Section \ref{sec_illustrations}. No issues of convergence are observed in either case. However, the mixing of the chains for the SIPP data is slower compared to the Durham and SDS data sets. 

\end{document}